\documentclass[11pt]{article}
\pdfoutput=1
\usepackage{jheppub,amsmath,amssymb}
\usepackage{bm}
\usepackage[table]{xcolor}
\usepackage{amsmath}
\usepackage{amssymb}
\usepackage{bbold}
\usepackage{graphics}
\usepackage{tikz}
\usepackage{todonotes}
\usetikzlibrary{arrows,positioning,decorations.markings,decorations.pathmorphing,calc}

\definecolor{hyperref}{RGB}{026,028,087}

\newcommand{\p}{{\bf p}}

\def\gsim{ \lower .75ex \hbox{$\sim$} \llap{\raise .27ex \hbox{$>$}} }
\def\lsim{ \lower .75ex \hbox{$\sim$} \llap{\raise .27ex \hbox{$<$}} }
\def\be{\begin{equation}}
\def\ee{\end{equation}}
\def\bea{\begin{eqnarray}}
\def\eea{\end{eqnarray}}

\newcommand{\ba}{\begin{array}}
\newcommand{\ea}{\end{array}}
\newcommand{\mn}{\mu\nu}

\newcommand{\commentout}[1]{}

\newcommand{\pa}{\partial}

\newcommand{\comment}[1]{}

\newcommand{\bs}{\begin{split}}

\newcommand{\eom}{{\it eom}}
\newcommand{\eoms}{{\it eoms}}
\newcommand{\dofs}{{\it dof's}}
\newcommand{\La}{\Lambda_3^3}
\newcommand{\dof}{{\it dof}}

\newcommand{\St}{{St\"uckelberg}}
\newcommand{\Os}{{Ostrogradsky}}

\def\ba{\begin{eqnarray}}
\def\ea{\end{eqnarray}}

\def\nn{\nonumber}

\def\d{\mathrm{d}}

\def\mupn{^\mu_{\, \nu}}

\def\({\left(}
\def\){\right)}

\definecolor{jn}{RGB}{10, 10, 200} 
\definecolor{js}{RGB}{204, 0, 0} 
\definecolor{pgf}{RGB}{10, 150, 10} 

\newcommand*{\mathcolor}{}
\def\mathcolor#1#{\mathcoloraux{#1}}
\newcommand*{\mathcoloraux}[3]{%
  \protect\leavevmode
  \begingroup
    \color#1{#2}#3%
  \endgroup
}
\newlength{\stheight}
\newcommand\textst[1][fu-grey]{
	\ifmmode\setlength{\stheight}{+1.0ex}
	\else\setlength{\stheight}{+0.5ex}
	\fi
	\bgroup\markoverwith{\textcolor{#1}{\rule[\the\stheight]{2pt}{1.0pt}}}\ULon
} 

\newcommand{\textins}[2][fu-grey]{
	\ifmmode\mathcolor{#1}{#2}
	\else\textcolor{#1}{#2}\@\,
	\fi
}

\def\D{\mathcal{D}}
\def\p{\partial}

\graphicspath{{./}}
\notoc

  \tikzstyle{vecArrow} = [thick, decoration={markings,mark=at position
   1 with {\arrow[semithick]{open triangle 60}}},
   double distance=1.4pt, shorten >= 5.5pt,
   preaction = {decorate},
   postaction = {draw,line width=1.4pt, white,shorten >= 4.5pt}]

\begin{document}

\title{The decoupling limit of Multi-Gravity: \\ Multi-Galileons, Dualities and More}

\author[a]{Johannes Noller}
\author[b]{, James H.C. Scargill}

\affiliation[a]{Astrophysics, University of Oxford, DWB, Keble Road, Oxford, OX1 3RH, UK} 
\affiliation[b]{Theoretical Physics, University of Oxford, DWB, Keble Road, Oxford, OX1 3NP, UK} 

\emailAdd{noller@physics.ox.ac.uk}
\emailAdd{james.scargill@physics.ox.ac.uk}

\abstract{
In this paper we investigate the decoupling limit of a particular class of multi-gravity theories, i.e. of theories of interacting spin-2 fields. We explicitly compute the interactions of helicity-0 modes in this limit, showing that they take on the form of multi-Galileons and dual forms. In the process we extend the recently discovered Galileon dualities, deriving a set of new multi-Galileon dualities. These are also intrinsically connected to healthy, but higher-derivative, multi-scalar field theories akin to `beyond Horndeski' models. 
}


\maketitle
\newpage

\setcounter{tocdepth}{2}

\tableofcontents

\section{Introduction}\label{sec-intro}

Investigating the decoupling limit of massive gravity-related theories (see \cite{Hinterbichler:2011tt, deRham:2014zqa} for reviews) has turned out to be a very fruitful exercise over the past few years. The decoupling limit, by which we mean the scaling limit of these theories isolating non-linear interactions between physical degrees of freedom (\dofs) suppressed by the smallest energy scale, reveals these \dofs{} and their low-energy interactions as well as strong coupling scales for the theories in question. Following in the footsteps of work investigating the decoupling limit of ghost-free (dRGT) Massive Gravity 
(see \cite{deRham:2010ik,deRham:2010kj,Hassan:2011hr} for the model itself and \cite{deRham:2010ik,Tasinato:2012ze,Ondo:2013wka,Gabadadze:2013ria,strong} and references therein for its decoupling limit interactions) and (Hassan-Rosen) Bigravity (see \cite{Hassan:2011tf,Hassan:2011zd,Hassan:2011ea} for the model itself and \cite{dualdiff} for investigations of its decoupling limit), we here probe the decoupling limit of (Hinterbichler-Rosen) Multi-Gravity theories \cite{Hinterbichler:2012cn} for the first time. For constraint analyses relevant for these models see \cite{Hinterbichler:2012cn,Deffayet:2012zc} and for related multi-gravity work also see \cite{Khosravi:2011zi,Nomura:2012xr,Hassan:2012wt,Tamanini:2013xia,cycles}.\footnote{Also, for a discussion of some of the as yet unresolved issues in the general field of massive gravity(-ies) and whether they are problematic for the field see e.g. \cite{Deser:2014fta} and the counterarguments in \cite{deRham:2014zqa}.}

One of the ways in which investigations of the decoupling limit have been very fruitful is by uncovering a set of dualities non-trivially relating (seemingly different, but ultimately physically equivalent) scalar \cite{dualdiff,galdual,gengaldual} and vector \cite{gengaldual} field theories via non-local, invertible field re-definitions. In particular these dualities can sometimes be useful in relating strongly to weakly coupled models. Scalar field dualities have been uncovered in this way by investigating helicity-0 mode interactions in the decoupling limit of Bigravity \cite{dualdiff}. These are described by Galileon interactions \cite{galileon} (also see \cite{Fairlie:1991qe,Horndeski:1974wa}) for the helicity-0 scalar \dof{} (as is also the case for massive gravity)  and the dualities relating these interactions can also be understood as abstracted duality transformations \cite{galdual} resulting from transformations on the coset space of $GAL(d, 1)/SO(d - 1, 1)$ \cite{Kampf:2014rka}.  The duality can also be extended to `generalized galileons' \cite{gengaldual}. 

The existence of these dualities is deeply linked to the non-uniqueness of the way in which gauge \dofs{} can be added to the theory in order to restore full diffeomorphism invariance - a feature that is made explicit in the link field formulation for Multi-Gravity theories \cite{ArkaniHamed:2001ca,ArkaniHamed:2002sp,ArkaniHamed:2003vb,Schwartz:2003vj,stuck}.
Using this formulation and the decoupling limit interactions for classes of Multi-Gravity models, in this paper we extend the (single) field Galileon dualities hitherto discovered to Multi-Galileon dualities and show how they can be understood both as inherited from diffeomorphism invariance and as abstracted duality transformations. 
\\

{\it Outline:} This paper is organised as follows. In section \ref{sec-dec} we review decoupling limit interactions for massive and bigravity and extend this to a number of ghost-free multi-gravity setups. In section \ref{sec-dual} we then discuss Galileon dualities in two ways: Firstly as a direct consequence of (diffeomorphism) gauge invariance in ghost-free Bigravity models and secondly as an abstract duality transformation. We then use the insight gained in this way to derive a set of multi-galileon dualities in section  \ref{sec-multidual}, where we also show how this can be used to demonstrate the healthiness of interactions between helicity-0 in multi-gravity models. Finally we discuss how matter can be coupled in multi-gravity theories and how this relates to the duality picture in section \ref{sec-matter}, before concluding in section \ref{sec-conc}.
\\

{\it Conventions}: Throughout this paper we use the following conventions. $D$ refers to the number of spacetime dimensions and we use Greek letters $\mu, \nu, \ldots$ as well as lower case Latin letters $a,b,\ldots$ to denote spacetime indices, which are raised and lowered with the Minkowski metric $\eta_{\mu\nu}$ unless explicitly stated otherwise. We write the completely anti-symmetric epsilon symbol as $\epsilon$ and define it such that $\epsilon_{012\cdots D}=1$ regardless of the signature of the metric or the position (up/down) of indices (hence $\epsilon^{012\cdots D}=\epsilon_{012\cdots D}=1$).

\section{Degrees of freedom, equations of motion and the absence of ghosts}\label{subsec-dof}

\subsection{Degrees of freedom and diffeomorphism invariance}

General relativity, a theory of a massless spin-2 particle, propagates two \dofs{} (around flat space and in the absence of a coupling to matter). A massive spin-2 field on the other hand propagates 5 \dofs{}. When building a massive gravity action, however, it is typically not obvious exactly how many \dofs{} propagate and what their dynamics is. The \St{} trick is useful in this circumstance, mapping a given action into another dynamically equivalent action, which contains more fields and more symmetry. In essence the \St{} trick is therefore the inverse of gauge-fixing an action, taking an action without a given symmetry and restoring it via the addition of gauge \dofs. For a more comprehensive review see \cite{stuck} - here we will be content to only give the bare recipe for performing the \St{} procedure.\footnote{Also see \cite{Gao:2014ula} for recent work on `covariant \St{} analyses'.} 
\\

{\bf Massive Gravity}.
Consider a generic local and Lorentz-invariant theory of massive gravity formulated in terms of a `metric' $g$ and a non-dynamical, flat background metric $\eta$
\be
{\cal S}_1 = M_{Pl}^{D-2} \int d^D x \sqrt{-g} R +  m^2 M_{Pl}^{D-2} \int d^D x \sqrt{-g} V\left(g^{-1} \eta \right).
\ee
The potential $V\left(g^{-1} \eta \right)$ breaks diffeomorphism invariance, which is just to say that it leads to the propagation of additional \dofs. We may restore full diffeomorphism invariance by performing either of the two following \St{} replacements patterned after the symmetry we are restoring and at the expense of introducing extra (gauge) fields
\bea
g_{\mu\nu} &\to & \pa_\mu Y^\alpha \pa_\nu Y^\beta g_{\alpha\beta}\left[Y(x)\right], \label{St1} \\
\eta_{\mu\nu} &\to & \pa_\mu \tilde Y^\alpha \pa_\nu \tilde Y^\beta \eta_{\alpha\beta}. \label{St2}
\eea
$Y^\alpha$ and $\tilde Y^\alpha$ are the \St{} fields encoding the extra gauge field content. That we have a choice of how to introduce these fields is a simple consequence of the fact that there is no unique procedure to introduce the redundancy associated with gauge fields when adding these fields to the action in the process of restoring diffeomorphism symmetry. 
Since in massive gravity we have a non-dynamical metric $\eta$, the result of using \eqref{St1} is an object which is invariant under the restored single copy of diffeomorphism invariance, rather than a tensor and so is more complicated to analyse than the result of using \eqref{St2} -- for details see \cite{Hinterbichler:2011tt}. We can now expand the \St{} fields via 
\bea
Y^\mu &\to & x^\mu + B^\mu + \pa^\mu \phi, \nn \\
\tilde Y^\mu &\to & x^\mu + A^\mu + \pa^\mu \pi,
\eea
respectively. $Y^\mu = x^\mu$ and $\tilde Y^\mu = x^\mu$ correspond to fixing the gauge fields to `unitary gauge'. In expanding the \St{} fields we have also introduced an additional $U(1)$ symmetry (effectively performing a second \St{} trick) with the associated fields $\pi$ and $\phi$ respectively. 
In the $\Lambda_3$ decoupling limit $A/B$ will describe the helicity-1 modes and $\pi/\phi$ the helicity-0 modes of the spin-2 fields in question. An important point is that the replacement \eqref{St2} is finite order in $A$ and $\pi$, whereas \eqref{St1} is not due to the dependence on $g_{\alpha\beta}\left[Y(x)\right]$ (a straightforward way to see this is to Taylor-expand $g_{\alpha\beta}\left[Y(x)\right]$ - see \cite{Hinterbichler:2011tt}). This makes identifying the interactions between different helicity modes very straightforward when \eqref{St2} is used to restore diffeomorphism invariance in the action.\footnote{Note that there is of course also a choice as to what symmetry one would like to restore, if any - see e.g. \cite{Bonifacio:2015rea} where instead of full diffeomorphisms, transverse diffeomorphisms and a Weyl symmetry are restored.} 
\\

{\bf Bi- and Multi-Gravity}.
Now consider a bi- or general multi-gravity model of the following type
\be
{\cal S}_2 = \sum_{i=1}^N M_{Pl}^{D-2} \int d^D x \sqrt{-g_{(i)}} R\left[g_{(i)}\right] +  m^2 M_{Pl}^{D-2} \int d^D x \sqrt{-g_{(1)}} V\left(g_{(1)}, \ldots, g_{(N)}) \right), \label{multiS}
\ee
where we have $N$ dynamical spin-2 fields described by $g_{\mu\nu}^{(i)}$, where bracketed indices $(i)$ are label indices only summed over if there is an explicit sum in what follows (i.e. there is no Einstein convention for label indices throughout this paper). The potential is a Lorentz-scalar built from the different $g_{(i)}$ and their inverses and all $N$ fields are dynamical in contrast to the massive gravity model above, which was really a bigravity theory with one non-dynamical (fixed) field $\eta$.\footnote{We have also assumed that the Planck masses for all spin-2 fields are the same and that the kinetic sector is solely made up from Einstein-Hilbert kinetic terms - for details on more complex kinetic sector interactions see \cite{Hinterbichler:2013eza,Folkerts:2011ev,Kimura:2013ika,deRham:2013tfa,
Gao:2014jja,kinetic}.}
Note that the choice of metric determinant in front of the potential can be absorbed into the definition of the potential and is therefore arbitrary. 
Let us focus on the bigravity version of \eqref{multiS}. The potential $V$ now generically breaks the two copies of diffeomorphism invariance present at the level of kinetic interactions for $g_{(1)}$ and $g_{(2)}$ down to the diagonal subgroup. The broken copy can again be restored via either of the two following replacements
\begin{align}
g_{\mu\nu}^{(1)} &\to  \pa_\mu Y^\alpha \pa_\nu Y^\beta g_{\alpha\beta}^{(1)}\left[ Y(x)\right] &\text{or}&
&g_{\mu\nu}^{(2)} &\to  \pa_\mu \tilde Y^\alpha \pa_\nu \tilde Y^\beta g_{\alpha\beta}^{(2)}\left[ \tilde Y(x) \right], 
\end{align}
where $Y$ and $\tilde Y$ can be expanded as before. Note that we no longer have the luxury of being able to introduce the \St{} fields via a non-dynamical field $\eta$ as before for theories of the type \eqref{multiS}, so these replacements are infinite order in the \St{} fields and in derivatives. As we shall see the fact that we can choose between two such replacements and hence two equivalent ways of capturing the helicity-0 mode ($\pi$ and $\phi$) here is intrinsically linked to the existence of Galileon dualities.

\begin{figure}[tp]
\centering
\begin{tikzpicture}[-,>=stealth',shorten >=0pt,auto,node distance=2cm,
  thick,main node/.style={circle,fill=blue!10,draw,font=\sffamily\large\bfseries},arrow line/.style={thick,-},barrow line/.style={thick,->},no node/.style={plain},rect node/.style={rectangle,fill=blue!10,draw,font=\sffamily\large\bfseries},red node/.style={rectangle,fill=red!10,draw,font=\sffamily\large\bfseries},green node/.style={circle,fill=green!20,draw,font=\sffamily\large\bfseries},yellow node/.style={rectangle,fill=yellow!20,draw,font=\sffamily\large\bfseries}]

 \node[circle,scale=1.3,fill=blue!30,line width=.5mm,draw=black,double](100){};
  \node[circle,scale=1.3,fill=white!30,line width=.5mm,draw=black,double] (101) [right of=100] {};
 
  \node[circle,scale=1.3,fill=white!30,line width=.5mm,draw=black,double] (2)  [right=4cm of 101]{};
   \node[circle,scale=1.3,fill=white!30,line width=.5mm,draw=black,double] (1) [left=1cm of 2] {};
  \node[circle,scale=1.3,fill=white!30,line width=.5mm,draw=black,double] (3) [right=1cm of 2] {};
   
    \node[circle,scale=1.3,fill=blue!30,line width=.5mm,draw=black,double] (5)  [right=5cm of 2]{};
  \node[circle,scale=1.3,fill=white!30,line width=.5mm,draw=black,double] (7) [below left=1cm of 5] {};
   \node[circle,scale=1.3,fill=white!30,line width=.5mm,draw=black,double] (8) [below right=1cm of 5] {};
   
     \node[draw=none,fill=none](80)[above left=1.5cm of 5]{};
       \node[draw=none,fill=none](81)[above right=1.5cm of 5]{};
        \node[draw=none,fill=none](82)[above=1.5cm of 5]{};
         \node[draw=none,fill=none](83)[right=0.7cm of 100]{};

  \path[every node/.style={font=\sffamily\small}]
  (100) edge node {} (101)
    (1) edge node {} (2)
     (3) edge node  {} (2)
      (7) edge node [above left] {} (5)
       (8) edge node [above right] {} (5);


\draw[-,dashed] (5) to (80);
\draw[-,dashed] (5) to (81);
\draw[-,dashed] (5) to (82);
    
\draw [<->] ($(8)+(0.3,0.2)$) 
    arc (-40:220:1.83);

   \node[draw=none,fill=none](92)[below of=83]{(a) Bigravity};
   
   \node[draw=none,fill=none](93)[below of=2]{(b) (Trimetric) line theory};
   
   \node[draw=none,fill=none](94)[below of=5]{(c) Star theory};
\end{tikzpicture}
\caption{Theory graph representations of different types of Multi-Gravity theories. Each node represents a metric $g_{(i)}$ and a double-circled node denotes the presents of an Einstein-Hilbert kinetic interaction $\sqrt{-g_{(i)}}R[g_{(i)}]$. Lines connecting two nodes are bigravity-like interaction terms between the two fields associated with those nodes (note that we do not include interactions between more than two metrics in this paper). Shaded nodes are coupled to matter (minimally in the cases shown here). See \cite{kinetic} for further details. The graphs shown depict: a) Hassan-Rosen Bigravity. b) A particular trimetric `line theory' as considered in sections \ref{subsec-multi} and \ref{sec-multidual}, with no coupling to matter specified. c) A `star theory' as considered in section \ref{sec-matter}, where several outer nodes are connected to a single central node, but not to each other. In the particular case shown here, only the central node is (minimally) coupled to matter.} \label{fig-interactions}
\end{figure}
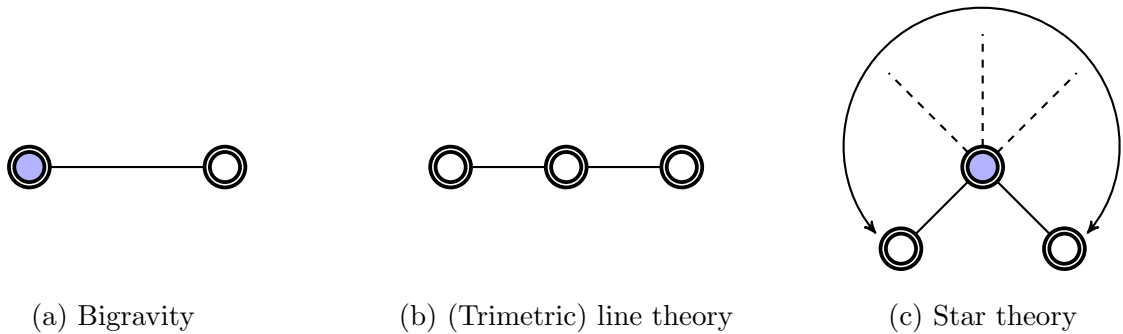

\subsection{Higher-derivative equations of motion} \label{subsec-higherDeoms}

Having performed the \St{} replacement discussed above one can read off the interactions of the different helicity modes and their interaction scales. The first check one would typically like to carry out, is to see whether the interactions found for the model under consideration are consistent - in particular: whether they are free of ghost-like instabilities. With an eye on the multi-gravity models we will consider in more depth later on, let us focus on the interaction between two scalars $\pi$ and $\phi$ at some scale $\Lambda$ (the scalars mimic helicity-0 modes and $\Lambda$ mimics the decoupling limit scale). When can we say these interactions are healthy (at least up to the scale $\Lambda$)?
\\

{\bf A multi-galileon}. Let us start with a familiar Lagrangian, namely that of a particular cubic multi-galileon (cf. \cite{Padilla:2010ir,Deffayet:2010zh,Hinterbichler:2010xn}).
\be
{\cal L}_1 = \frac{1}{2}\pi\Box\pi + \frac{1}{2}\phi\Box\phi + \frac{1}{\Lambda}\pi \left(\partial_{b}\partial_{a}\phi \partial^{b}\partial^{a}\phi - \left(\Box\phi\right)^2  \right).
\ee
The equations of motion (\eoms) for this Lagrangian are
\bea
{\cal E}_{\phi} &=&  \Box\phi + \frac{2}{\Lambda} \left(\partial_{b}\partial_{a}\pi \partial^{b}\partial^{a}\phi -\Box\phi \Box\pi \right) = 0,
\nn \\
{\cal E}_{\pi} &=& \Box\pi + \frac{1}{\Lambda} \left(\partial_{b}\partial_{a}\phi \partial^{b}\partial^{a}\phi -  \left(\Box\phi\right)^2 \right) = 0.
\eea
These \eoms{} are manifestly second order and the linear kinetic term has the correct sign. This is a quick diagnostic to show that the system is free of (Ostrogradsky) ghost-like instabilities. If the \eoms{} have dependence on higher-order derivatives, in the absence of additional constraints/degeneracies this will lead to propagation of an extra ghostly \dof{} commonly called an Ostrogradsky ghost \cite{Ostro,Woodard:2006nt}. 
\\

{\bf A sick theory}. A quick example of such a ghostly, higher order in derivatives Lagrangian is given by a small modification of ${\cal L}_1$, where we have broken the antisymmetric structure of the cubic interaction term 
\be
{\cal L}_2 = \frac{1}{2}\pi\Box\pi + \frac{1}{2}\phi\Box\phi + \frac{1}{\Lambda}\pi \Box \pi \Box\phi.
\ee
The \eoms{} now are
\bea
{\cal E}_{\phi} &=&  \Box\phi + \frac{1}{\Lambda} \left( (\Box\pi)^2 + 2 \partial^{a}\pi \partial_{a}\Box\pi + \pi \Box^2 \pi \right) = 0, \nn \\
{\cal E}_{\pi} &=& \Box\pi + \frac{1}{\Lambda} \left(2 \Box\phi \Box \pi + 2 \partial^{a}\pi \partial_{a}\Box\phi + \pi \Box^2 \phi\right) = 0,
\eea
and they are explicitly higher-order. In particular, the first \eom{} depends on up to fourth derivatives of $\pi$, whereas the second \eom{} depends on up to fourth derivatives of $\phi$. Consequently more than two initial conditions need to be specified per field and the extra \dof{} encoded in these additional initial conditions is an Ostrogradsky ghost. Note that, if an invertible, non-linear field re-definition can be found, which removes interactions at this order in the fields, the addition of higher order (in the fields) interactions to the Lagrangian can prevent the appearance of a ghost at the scale $\Lambda$.
\\

{\bf Healthy higher-derivative \eoms}. Finally let us consider a case, where the \eoms{} appear higher-order, yet are ghost-free due to the existence of an extra constraint. The following is a straightforward example of this situation, which we have lifted from \cite{deRham:2011qq}
\be
{\cal L}_3 = \frac{1}{2}\pi\Box\pi + \frac{1}{2}\phi\Box\phi + \frac{1}{\Lambda^5}\Box \pi \pa_{a} \pa_b \phi \pa^{a} \pa^b \phi + \frac{1}{2 \Lambda^{10}}\pa_a \pa_b \phi \pa^a \pa^b \phi  \Box(\pa_c \pa_d \phi \pa^c \pa^d \phi)
\ee
The equations of motion here are
\bea
{\cal E}_\phi &=& \Box\phi + \frac{2}{\Lambda^5} \pa_a \pa_b \left(\pa^a \pa^b \phi \left[\Box\pi + \frac{1}{\Lambda^5}\Box\left(\pa_d \pa_e \phi \pa^d \pa^e \phi \right)  \right]\right) = 0, \nn \\
{\cal E}_\pi &=& \Box\pi + \frac{1}{\Lambda^5}\Box(\pa_d \pa_e \phi \pa^d \pa^e \phi) = 0,
\eea
which appear to be higher order in derivatives. However, note the explicit dependence of the first \eom{} on the second, which is the way in which the extra constraint manifests itself here. By writing $\hat {\cal E}_\phi = {\cal E}_\phi - 2/\Lambda^5 \pa_a \pa_b \left( \pa^a\pa^b\phi {\cal E}_\pi\right)$ and $\hat {\cal E}_\pi = {\cal E}_\pi$, we can instead write the \eoms{} for $\phi$ and $\pi$ as
\bea
\hat {\cal E}_\phi \left[\phi^{II}\right]&=& \Box \phi = 0, \nn \\
\hat {\cal E}_\pi \left[\pi^{II}, \phi^{II}, \phi^{III}, \phi^{IV} \right]&=& \Box\pi + \frac{1}{\Lambda^5}\Box(\pa_d \pa_e \phi \pa^d \pa^e \phi) = 0,
\eea
where we have made the dependence on derivatives explicit, i.e. $\hat {\cal E}_\pi \left[\pi^{II}, \ldots \right]$ denotes that this \eom{} depends on second derivatives acting on $\pi$ and so on. It is then clear that $\phi$ can be solved for in terms of just two initial conditions. $\hat {\cal E}_\pi$ is higher-derivative in nature, but the higher-derivative dependence is restricted to $\phi$, for which we have already solved. So $\pi$ can also be solved for in terms of two initial conditions and the whole system is secretly second-order with a well-defined Cauchy problem. Schematically we have derived a final set of \eoms{}, $\tilde {\cal E}_\pi, \tilde {\cal E}_\phi$, where $\tilde {\cal E}_\phi = \hat {\cal E}_\phi$ and $\tilde {\cal E}_\pi$ was obtained by substituting the solution of $\hat {\cal E}_\phi$ into $\hat {\cal E}_\pi$. Consequently we have two final \eoms{} $\tilde {\cal E}_\phi \left[\phi^{II}\right], \tilde {\cal E}_\pi \left[\pi^{II},\phi^{II}\right]$. The same conclusion could have been reached by performing an explicit constraint analysis for this model in the Hamiltonian picture. In general, if there is any linear combination of the \eoms{} of a system and derivatives of the \eoms{} that explicitly allows all the fields to be solved in terms of two initial conditions for each field (here we are assuming we are dealing with scalars, obviously this count is modified for other tensors), then we have a second-order system free from Ostrogradsky ghosts. Note that the models proposed in \cite{Gabadadze:2012tr} as well as the recently proposed `beyond Horndeski' theories \cite{Zumalacarregui:2013pma,Gleyzes:2014dya} are precisely of this type as well (albeit being theories of a scalar and a metric tensor, instead of multiple scalars as discussed here, in the case of `beyond Horndeski' theories).

\subsection{Some notation}\label{subsec-not}

Let us quickly summarise some conventions used throughout the remainder of this paper. We will reserve the letters $\pi$ and $\phi$ for \St{} scalars. In order to keep notation concise, we will typically simply denote partial derivatives acting on these scalars by indices, i.e.
\be
\pi_{a_1 \ldots a_n}^{b_1 \ldots b_n} \equiv \pa_{a_1} \ldots \pa_{a_n} \pa^{b_1} \ldots \pa^{b_n} \pi
\ee
Both Latin and Greek letters denote space-time indices unless stated otherwise. Bracketed indices as in $\pi_{(1)}$ are label indices, e.g. labelling different helicity-0 modes in multi-gravity setups. Label indices are never summed over. In order to connect with notation used in the literature, we will sometimes denote $\pi_a^b$ by $\Pi_a^b$, 
\ba
\Pi\mupn(x)=\frac{1}{\Lambda^3}\eta^{\mu\alpha}\partial_\alpha\partial_{\nu} \pi(x)\,,
\ea
i.e. we will reserve $\Pi$ to denote this particular two derivative function of $\pi$, especially when using index-free notation - see below. $\Lambda$ is a mass scale, which in the 4D massive gravity context will turn out to be $\Lambda_3 = (m^2 M_{Pl})^{1/3}$. For a scalar $\pi$, at each order $n$, there is a unique total derivative combination given by 
\be \label{TD}
{\cal L}^{\text{TD}}_{(n)}(\Pi) = \delta^{\alpha_1 \ldots \alpha_n}_{[\beta_1 \ldots \beta_n]} \pi^{\beta_1}_{\alpha_1} \ldots \pi^{\beta_n}_{\alpha_n},
\ee
where we have defined a tensor $\delta^{\alpha_1 \ldots \alpha_n}_{[\beta_1 \ldots \beta_n]}$ separately anti-symmetric in its indices $\alpha_1 \ldots \alpha_n$ and $\beta_1 \ldots \beta_n$ in terms of the totally antisymmetric tensor $\varepsilon$ via 
\be \label{delta-def}
\delta^{\alpha_1 \ldots \alpha_n}_{[\beta_1 \ldots \beta_n]}  \equiv \frac{1}{(D-n)!}\varepsilon^{\alpha_1 \ldots \alpha_n \lambda_1 \ldots \lambda_{D-n}}
\varepsilon_{\beta_1 \ldots \beta_n \lambda_1 \ldots \lambda_{D-n}}.
\ee
Equivalently to ${\cal L}^{\text{TD}}_{(n)}(\Pi)$, and in order to connect with the notation used by \cite{galdual,dualdiff,gengaldual}, we may also define the characteristic polynomials $U_{(n)}$ of a matrix $M$ in the following index-free way
\be\label{UN-def}
U_{(n)}(M) \equiv \epsilon \epsilon M^n \eta^{D-n},
\ee
where $U_{(n)}(\Pi) = (D-n)! {\cal L}^{\text{TD}}_{(n)}(\Pi)$. To give an explicit example, for $U_{(2)}$ in 4D this means
\bea
U_{(2)}(\Pi) &=& \epsilon^{abcd} \epsilon^{\mu\nu\rho\sigma} \Pi_{\mu a} \Pi_{\mu b} \eta_{\rho c} \eta_{\sigma d} \nn \\
&=& 2!\delta^{\alpha_1 \alpha_2 \alpha_3 \alpha_4}_{[\beta_1 \beta_2 \beta_3 \beta_4]}\Pi_{\alpha_1}^{\beta_1} \Pi_{\alpha_2}^{\beta_2} \delta_{\alpha_3}^{\beta_3} \delta_{\alpha_4}^{\beta_4} \nn \\
&=& 2{\cal L}^{\text{TD}}_{(2)}(\Pi).
\eea
In addition we can define the tensor $X_{(n)}^{\mn}(\Pi)$, which is the unique symmetric and identically conserved function of $\Pi$ at a given order $n$
\be \label{X-def}
X_{(n)}{}^{\mu}{}_\nu(\Pi) = \delta^{\mu \mu_1 \hdots \mu_n}_{[\nu \nu_1 \hdots \nu_n]} \Pi^{\nu_1}_{\mu_1} \hdots \Pi^{\nu_n}_{\mu_n}
\ee
We will also find it useful to define the related tensors $X^{\mn}(\Pi)$ and $\tilde X^{\mn}(\Pi)$, which mix all orders in $n$ 
\bea
X^{\mn}(\Pi) &=& -\frac{1}{2}\sum_{n=0}^{D-1} \frac{\hat \beta_n}{(D-1-n)!n!} \epsilon^{\mu\ldots}\epsilon^{\nu\ldots} (\eta + \Pi)^n \eta^{D-1-n},\nn \\
\tilde X^{\mn}(\Pi) &=& -\frac{1}{2}\sum_{m=1}^{D} \frac{\hat \beta_m}{(D-m)!(m-1)!} \epsilon^{\mu\ldots}\epsilon^{\nu\ldots} (\eta + \Pi)^{D-m} \eta^{m-1},
\eea
where the $\hat \beta_n$ are constant coefficients. Note that $X$ and $\tilde X$ can straightforwardly be mapped into each another with the replacement $m = D-n$ - they are essentially the same object with remapped coefficients\footnote{Note that our $\tilde X$ is essentially the same object as $\tilde Y$ defined by \cite{dualdiff}. We choose our notation in order to emphasise that $X$ and $\tilde X$ denote identical interactions up to constant coefficients and since this definition will be rather useful in the general multi-gravity treatment of section \ref{sec-multidual}.}. Characteristic polynomials , the `$X$-tensors' defined above and total derivative combinations at a given order are related by
\bea
\eta_{\mn} X^{\mn}_{(i)}(\Pi) &=& \frac{1}{(D-1-i)!}U_{(i)}(\Pi), \nn \\
\eta_{\mn}X^{\mn}(\Pi) &=& -\frac{1}{2}\sum_{n=0}^{D} \frac{\hat \beta_n}{(D-1-n)!n!}U_{(n)}(\eta +\Pi), \nn \\
\eta_{\mn}\tilde X^{\mn}(\Pi) &=& -\frac{1}{2}\sum_{m=0}^{D} \frac{\hat \beta_m}{(D-m)!(m-1)!}U_{(m)}(\eta +\Pi), \nn \\
X^{\mn}_{(n)}(\Pi) &=& \frac{1}{n+1} \frac{\delta}{\delta\Pi_{\mn}} {\cal L}_{n+1}^{\rm TD}(\Pi).
\eea

\section{The decoupling limit of Massive, Bi-, and Multi-gravity}\label{sec-dec}

In this section we will review the decoupling limit interactions for massive and bigravity and derive  analogous results for a class of multi-gravity models. 

\subsection{Massive gravity}\label{subsec-dRGT}

The ghost-free dRGT massive gravity model \cite{deRham:2010ik,deRham:2010kj,Hassan:2011hr} can be written down as
\be
\label{dRGT-act} 
S=\int d^D x \bigg[ M_{\rm Pl}^{D-2} \sqrt{- g}\,R[g] + m^2 M_{\text{Pl}}^{D-2} \sqrt{- g}\,\sum_{n=0}^{D} \beta_n\,
  U_n(\sqrt{g^{-1} \eta})\bigg] ,
\ee
where the $\beta_n$ are constant coefficients, $U_n$ denotes an elementary symmetric polynomial of order $n$ and (in index-free matrix notation) $\sqrt{g^{-1} \eta} = M$, where $M$ is any real matrix satisfying $M^2 = g^{-1} \eta$.\footnote{Note that the existence of such a real square root is a non-trivial requirement, related to the symmetric vielbein condition \cite{Deffayet:2012zc}.}
We now perform the \St{} procedure, restoring full diffeomorphism invariance as shown above (we here choose \eqref{St2}) and making interactions of the different helicity modes explicit in the process. Expanding the metric around a flat background, $g_{\mu\nu} = \eta_{\mu\nu} + h_{\mu\nu}$, one then finds that interactions involving $n_\pi$ fields $\pi$, $n_A$ fields $A_{\mu}$ and $n_h$ fields $h_{\mu\nu}$, carry an interaction scale
\begin{equation} \label{intscales}
\Lambda_{\lambda}=\left(M_{\rm Pl}^{\frac{D-2}{2}} \; m^{\lambda-1}\right)^{1/\lambda}= m\left( \frac{M_{\rm Pl}^{\frac{D-2}{2}}}{m} \right)^{1/ \lambda} , \ \ \ \lambda=\frac{3n_\pi+2n_A+n_h-4}{n_\pi+n_A+n_h-2},
\end{equation}
The least suppressed interactions that survive in the dRGT model carry a scale $\Lambda_3$.\footnote{In 4D this is a mass scale.} We can isolate interactions at this scale by taking the $\Lambda_3$ decoupling limit \footnote{Note that in this limit the dynamics of the helicity-0 mode is captured by the \St{} scalar $\pi$, the helicity-1 modes are captured by the $A_\mu$ and the helicity-2 modes are captured by $h_{\mu\nu}$, whereas beyond the decoupling limit (i.e. for more suppressed interaction terms), this is not the case \cite{deRham:2011rn}. This is beyond the scope of this paper, however, and we restrict ourselves to a discussion of the decoupling limit phenomenology throughout.}
\begin{align}
m &\rightarrow 0, &M_{\rm Pl} &\rightarrow \infty, &\Lambda_3 &\text{ fixed}.
\end{align}
The helicity-1 mode $A$ does not appear linearly in the action, so we can consistently set it to zero, which we will do for the remainder of this paper.\footnote{In fact, as long as matter is minimally coupled to the metric, $A$ never gets sourced. For details on helicity-1 interactions in massive gravity see \cite{Ondo:2013wka,Gabadadze:2013ria}.} 

In the $\Lambda_3$ decoupling limit the helicity-2 and helicity-0 modes are described by $h$ and $\pi$ respectively -- they are now governed by the following action
\bea  
S &=& \int d^Dx \left[ -\frac{1}{4} h_{\mu\nu}{\cal E}^{\mu\nu}_{\alpha\beta} h^{\alpha\beta} + \frac{\Lambda_3^3}{2} \sum_{n=1}^{D-1} \alpha_{(n)} h^{\mn} X_{\mn}^{(n)}(\Pi) \right], \nn \\
&=& \int d^Dx  \left[-\frac{1}{4} h_{\mu\nu}{\cal E}^{\mu\nu}_{\alpha\beta} h^{\alpha\beta} + \frac{\Lambda_3^3}{2} h^{\mn} X_{\mn}(\Pi)\right], \label{massivedec}
\eea
where the $\alpha_{(i)}$ are constant coefficients and we have ignored the coupling to matter.\footnote{Note that $ X_{\mn}^{(4)}(\pi)$ and higher orders in $X$ vanish identically in 4D. So the $h\pi$ couplings shown in \eqref{massivedec} are all the non-vanishing $h^{\mn} X_{\mn}^{(n)}(\pi)$ terms in 4D.} We can now demix the interactions between $h$ and $\pi$ via the following field redefinitions
\bea
h_{\mn} &\to & h_{\mn} + \frac{2 \alpha_{(1)} \La}{D-2} \pi \eta_{\mn}, \label{massivedemix1} \\
h_{\mn} &\to & h_{\mn} - 2 \alpha_{(2)} \La \pi_\mu \pi_\nu.  \label{massivedemix2}
\eea
The first one, a linearised conformal transformation, demixes the different helicity modes at linear order, eliminating the $ h^{\mn} X_{\mn}^{(1)}(\pi)$ mixing. Subsequently applying the second field re-definition also eliminates the cubic $ h^{\mn} X_{\mn}^{(2)}(\pi)$ mixing - demixing at higher orders cannot be achieved with a local field re-definition. After these replacements, the helicity-0 self-interactions at cubic, quartic and quintic order (in 4D these are the only non-vanishing terms) are given by 
\bea
S_{(3)} &\sim & \int d^Dx \; \pi_a \pi^a \pi_b^b, \\
S_{(4)} &\sim & \int d^Dx \; \pi_a \pi^a \left(\pi_b^b \pi_c^c - \pi_b^c \pi_c^b \right), \\
S_{(5)} &\sim & \int d^Dx \; \pi_a \pi^a
\left(3 \pi^{d}{}_{d} \pi_{bc} \pi^{bc} -2 \pi_{d}{}^{c} \pi^{db} \pi_{bc} - \pi^{d}{}_{d} \pi^{b}{}_{b} \pi^{c}{}_{c}\right).
\eea
Note that $S_3$ and $S_4$ are already present after the first field redefinition \eqref{massivedemix1}, whereas $S_5$ only appears after \eqref{massivedemix2} has been performed.\footnote{This is also a quick way of seeing that there is only a two-parameter family of solutions present in the decoupling limit - a statement which is also true for the full theory.} The corresponding equations of motion are ${\cal E}_{(i)} = 0$, where 
\bea
{\cal E}_{(3)} &=& \pi_a^a \pi_b^b - \pi_a^b \pi_b^a = \frac{1}{(D-2)!}U_{(2)}(\Pi),  \\
{\cal E}_{(4)} &=& 3 \pi^{a}{}_{a} \pi_{bc} \pi^{bc} -2 \pi_{a}{}^{c} \pi^{ab} \pi_{bc} -  \pi^{a}{}_{a} \pi^{b}{}_{b} \pi^{c}{}_{c} =  \frac{1}{(D-3)!}U_{(3)}(\Pi), \\
{\cal E}_{(5)} &=&  8 \pi^{a}{}_{a} \pi_{b}{}^{d} \pi^{bc} \pi_{cd} -6 \pi_{a}{}^{c} \pi^{ab} \pi_{b}{}^{d} \pi_{cd} + 3 \pi_{ab} \pi^{ab} \pi_{cd} \pi^{cd} \\ \nn  &-& 6 \pi^{a}{}_{a} \pi^{b}{}_{b} \pi_{cd} \pi^{cd} +  \pi^{a}{}_{a} \pi^{b}{}_{b} \pi^{c}{}_{c} \pi^{d}{}_{d} = \frac{1}{(D-4)!}U_{(4)}(\Pi).
\eea
As is well-known, these terms are precisely the Galileon interactions \cite{galileon} and their corresponding \eoms{} (where we avoid any integration-by-parts ambiguities). Consequently we may write the action for the helicity-0 mode in the decoupling limit as
\bea
S = \int d^D x \sum_i c_{(i)} \pi U_n (\Pi).
\eea
Here all higher order equations of motion for $\pi$ vanish and we can in fact look at each order in the \eoms{} separately to confirm that they are (purely) 2nd order in derivatives of $\pi$ at all orders. Consequently $\pi$ does not give rise to any ghost-like (\Os) \dof.

\subsection{Bigravity}\label{subsec-HR}

The Hassan-Rosen bigravity model \cite{Hassan:2011tf,Hassan:2011zd,Hassan:2011ea}, essentially the extension of the dRGT massive gravity model in which both metrics are dynamical, is described by the following action
\be
\label{HR-act} 
S=\int d^Dx \bigg[ M_{\rm Pl}^{D-2} \sqrt{- g}\,R[g] + M_{\rm Pl}^{D-2} \sqrt{- f}\,R[f] + m^2 M_{\text{Pl}}^{D-2} \sqrt{- g}\,\sum_{n=0}^{D} \beta_n\,
  U_n(\sqrt{g^{-1} f})\bigg] ,
\ee
where we have assumed that the Planck masses for both metrics are identical - a generalisation is straightforward. As before we perform the \St{} replacement, ignore the helicity-1 modes for the time being and focus on the interactions of the helicity-2 and helicity-0 modes in the decoupling limit.\footnote{Since the two Planck masses are the same, this limit is the same as before, otherwise different decoupling limits are possible \cite{stuck}.} 
Due to a symmetry of the elementary symmetric polynomials $U_{(n)}$, namely $\sqrt{-g}U_{(n)}(\sqrt{g^{-1}f}) = \sqrt{-f}U_{(D-n)}(\sqrt{f^{-1}g})$, these can be written \cite{dualdiff}
\bea
S &=& \int \d^Dx \left[  -\frac{1}{4} h^{\mu \nu}_{(1)} \hat{\mathcal{E}}^{\alpha \beta}_{\mu\nu} h_{\alpha\beta}^{(1)} -\frac{1}{4} h^{\mu \nu}_{(2)} \hat{\mathcal{E}}^{\alpha \beta}_{\mu\nu} h_{\alpha\beta}^{(2)} + \frac{\Lambda_3^3}{2}h_{\mu\nu}^{(1)} X^{\mu\nu}(\pi) + \frac{\Lambda_3^3}{2} h_{\mu \nu}^{(2)} \tilde X^{\mu \nu}(\Sigma) \right],
\eea
where $\sigma$ is a field `dual' to $\pi$ and can be implicitly defined via\footnote{We can also define $\sigma$ via an implicit function $Z$ at this point \cite{dualdiff}
\begin{align}
Z^a(x^b + \frac{1}{\Lambda_3^{3}}\partial^b \pi(x)) &= x^a,
&Z^a(x) &= x^a + \frac{1}{\Lambda_3^3} \partial^a \sigma(x) \, .
\end{align}
We will discuss the origin and form of $\sigma$ in more detail in the next section.}
\be
x + \partial \sigma = (x + \partial \pi)^{-1}, \label{dual relation 1}
\ee
and in analogy to our definition of $\Pi$ we have also defined $\Sigma$ via $\Sigma \mupn(x)=\partial^{\mu}\partial_{\nu} \sigma( x)/\Lambda_3^{3}$. We can then iteratively solve for $\sigma$ in terms of $\pi$ and find (to quintic order in $\pi$)
\bea \label{sigma-itsolve}
\sigma =  &-& \pi + \frac{1}{2 \La} \pi_{a} \pi^{a} -  \frac{1}{2 \Lambda_3^6} \pi^{a} \pi^{b} \pi_{ab} + \frac{1}{2 \Lambda_3^9} \pi^{a} \pi^{b} \pi_{a}{}^{c} \pi_{bc} + \frac{1}{6 \Lambda_3^9} \pi^{a} \pi^{b} \pi^{c} \pi_{abc} \\ &-&  \frac{1}{2 \Lambda_3^{12}} \pi^{a} \pi^{b} \pi_{a}{}^{c} \pi_{b}{}^{d} \pi_{cd}  -  \frac{1}{2 \Lambda_3^{12}} \pi^{a} \pi^{b} \pi^{c} \pi_{a}{}^{d} \pi_{bcd} -  \frac{1}{24 \Lambda_3^{12}} \pi^{a} \pi^{b} \pi^{c} \pi^{d} \pi_{abcd} + {\cal O}(\pi^6). \nn
\eea
We will look into the relation between $\sigma$ and $\pi$ in more detail in section \ref{sec-dual} (in particular a full solution is given in \eqref{sigmatopi}), but looking ahead it is worth pointing out that $\sigma$ is related to $\pi$ via a non-local, but invertible, field redefinition \cite{galdual,dualdiff}. This is important because such a field redefinition preserves the number of \dofs{} (in this case the \dof{} described by $\pi$). 
\\

{\bf Diagonalising kinetic terms}.
We now demix the action at linear order (for full details see \cite{stuck}). There are a number of possible field re-definitions we may choose to do so, two of which are
\begin{align}
{\cal M}_1 &:& h_{\mn}^{(1)} &\to h_{\mn}^{(1)} + c_1 \pi \eta,  &h_{\mn}^{(2)} \to h_{\mn}^{(2)} + c_2 \sigma \eta, \nn \\
{\cal M}_2 &:& h_{\mn}^{(1)} &\to h_{\mn}^{(1)} + c_1 \pi \eta,  &h_{\mn}^{(2)} \to h_{\mn}^{(2)} + c_2 \pi \eta. \label{demixings}
\end{align}
As far as demixing at linear order goes, the difference between ${\cal M}_1$ and ${\cal M}_2$ is irrelevant, since to first order $\sigma(\pi) = -\pi$, but we will see that ${\cal M}_1$ makes the number of propagating \dofs{} and their interactions more explicit. When applying  ${\cal M}_1$ we obtain the following schematic action for the pure helicity-0 piece post-demixing (and still in the decoupling limit)
\bea \label{Sdemix1}
S_1 = \int d^D x \sum_i \left( c_{(i)} \pi U_n (\Pi) + d_{(i)} \sigma U_n (\Sigma) \right),
\eea
i.e. we have an explicitly second-order galileon-like term for $\pi$ and one for $\sigma$. However, $\sigma$ and $\pi$ describe the same \dofs{} and we should therefore really express the action in terms of just one of the two. What is not trivial to see is that the this action also only propagates a single scalar \dof{} when $\sigma$ is expressed in terms of $\pi$ or vice versa. Naively one may expect that the then higher derivative nature of the action would lead to the appearance of additional propagating \dofs.\footnote{In this subsection we will directly inspect the \eoms{} and deduce the number of propagating \dofs{} that way. Independently the bigravity duality picture as discovered by \cite{dualdiff,galdual} and discussed in section \ref{sec-dual} will also show independently that no additional \dof{} propagates due to the apparent higher-order nature of the action.}
When applying ${\cal M}_2$, on the other hand, we obtain 
\bea \label{Sdemix2}
S_2 = \int d^D x \sum_i \left( c_{(i)} \pi U_n (\Pi) + d_{(i)} \pi U_n (\Sigma) \right),
\eea
where it is now no longer clear that the second term describes a Galileon-like interaction and the true number of propagating \dofs{} is not obvious. 
\\

{\bf Equations of motion}.  
We now again move on to examine the \eom{} for $\pi$, where we directly express $\sigma$ in terms of $\pi$ via \eqref{sigma-itsolve}. We already know that the $\pi U_n (\Pi)$ piece in \eqref{Sdemix1} will contribute second-order Galileon terms to the \eom. In addition from the $\sigma U_n (\Sigma)$ in \eqref{Sdemix1} we obtain the following contributions to the \eom{} when varying with respect to $\pi$
\bea
{\cal E}_{(1)} &=& 0, \nn\\
{\cal E}_{(2)} &=& \hat\beta_1^{\text{}} \pi^{a}{}_{a},\nn\\
{\cal E}_{(3)} &=& \hat\beta_1^{\text{}} (\pi^{a}{}_{a} \pi^{b}{}_{b} - \pi_{ab} \pi^{ab}) + \hat\beta_2^{\text{}} (\tfrac{3}{2} \pi_{ab} \pi^{ab} -  \tfrac{3}{2} \pi^{a}{}_{a} \pi^{b}{}_{b}),\nn\\
{\cal E}_{(4)} &=& \beta_1^{\text{}} (\pi_{a}{}^{c} \pi^{ab} \pi_{bc} -  \tfrac{3}{2} \pi^{a}{}_{a} \pi_{bc} \pi^{bc} + \tfrac{1}{2} \pi^{a}{}_{a} \pi^{b}{}_{b} \pi^{c}{}_{c})  \nn
\\
&+& \hat\beta_2^{\text{}} (\tfrac{9}{2} \pi^{a}{}_{a} \pi_{bc} \pi^{bc} -  \tfrac{3}{2} \pi^{a}{}_{a} \pi^{b}{}_{b} \pi^{c}{}_{c} -3 \pi_{a}{}^{c} \pi^{ab} \pi_{bc}) \nn
\\
&+& \hat\beta_3^{\text{}} (2 \pi_{a}{}^{c} \pi^{ab} \pi_{bc} - 3 \pi^{a}{}_{a} \pi_{bc} \pi^{bc} + \pi^{a}{}_{a} \pi^{b}{}_{b} \pi^{c}{}_{c}), \label{firsteoms}
\eea
where ${\cal E}_{(i)}$ denotes the contribution to the \eom{} coming from the i-th order (in $\pi$) piece in the action. ${\cal E}_{(i)}$ is consequently $(i-1)$-th order in $\pi$. We have also omitted quintic order (this can be found in appendix \ref{appendix-quintic}) and set $\La = 1$ to avoid clutter. We can immediately see that all the contributions are second order in derivatives and this remains true at quintic order, i.e. the Galilean symmetry is manifest. Importantly, up to quintic order in the fields at least, there is therefore no \Os{} instability arising from helicity-0 interactions in ghost-free bigravity.

Suppose we had demixed differently with ${\cal M}_2$ and consequently obtained \eqref{Sdemix2} instead. The extra (not obviously of Galileon form) contribution to the \eom{} now comes from the $\pi U_n (\Sigma)$ piece in \eqref{Sdemix2}. This results in the following contribution when varying with respect to $\pi$ 
\bea
{\cal E}_{(1)} &=& 0,\nn\\
{\cal E}_{(2)} &=& \hat\beta_1^{\text{}} \pi^{a}{}_{a},\nn\\
{\cal E}_{(3)} &=&  \hat\beta_1^{\text{}} (\tfrac{1}{2} \pi^{a}{}_{a} \pi^{b}{}_{b} - \tfrac{1}{2} \pi_{ab} \pi^{ab}) + \hat\beta_2^{\text{}} (\tfrac{3}{2} \pi_{ab} \pi^{ab} -  \tfrac{3}{2} \pi^{a}{}_{a} \pi^{b}{}_{b}),\nn\\
{\cal E}_{(4)} &=& 
\hat\beta_1^{\text{}} (\tfrac{1}{2} \pi_{a}{}^{c} \pi^{ab} \pi_{bc} -  \tfrac{1}{4} \pi^{a}{}_{a} \pi_{bc} \pi^{bc} + \tfrac{1}{4} \pi^{a}{}_{a} \pi^{b}{}_{b} \pi^{c}{}_{c} \nn\\ &+& \pi^{a} \pi^{bc} \pi_{abc} + \tfrac{1}{2} \pi^{a} \pi^{b}{}_{b} \pi_{a}{}^{c}{}_{c} + \tfrac{1}{2} \pi^{a} \pi_{a}{}^{b} \pi_{b}{}^{c}{}_{c} + \tfrac{1}{2} \pi^{a} \pi^{b} \pi_{ab}{}^{c}{}_{c})\nn\\
&+& \hat\beta_2^{\text{}} (3 \pi^{a}{}_{a} \pi_{bc} \pi^{bc} -2 \pi_{a}{}^{c} \pi^{ab} \pi_{bc} -  \pi^{a}{}_{a} \pi^{b}{}_{b} \pi^{c}{}_{c}) \nn\\
&+& \hat\beta_3^{\text{}} (2 \pi_{a}{}^{c} \pi^{ab} \pi_{bc} - 3 \pi^{a}{}_{a} \pi_{bc} \pi^{bc} + \pi^{a}{}_{a} \pi^{b}{}_{b} \pi^{c}{}_{c}). \label{secondeoms}
\eea
Up to cubic order in $\pi$ and ignoring irrelevant overall coefficients these contributions to the \eom{} are identical to those derived from \eqref{Sdemix1}. However, at quartic order they become different and noticeably we now have a higher-derivative dependence (the second line of ${\cal E}_{(4)}$ contains all these terms). In particular these terms all only depend on $\hat\beta_1$. If we continue to quintic order (see appendix \ref{appendix-quintic}) we find that there is higher-derivative dependence via $\hat \beta_2$ now as well. This is a direct consequence of the fact that the two different demixing procedures agree up to linear order, but diverge at higher orders in the field. The higher derivative dependence of the second set of contributions \eqref{secondeoms} is inherited from the action's dependence on $\sigma$, cf. \eqref{sigma-itsolve}, which is non-locally related to $\pi$, i.e. infinite order in derivatives. The fact that the action now non-locally depends on $\pi$ means that \Os{} counting is not(!) appropriate for the full theory - the action is degenerate since one cannot isolate a highest-order-in-derivatives term \cite{Ostro,Woodard:2006nt}. The fact that the first set of contributions \eqref{firsteoms}, obtained via a different demixing procedure, is purely second-order is already a direct hint that a non-local, invertible field re-definition exists for $\pi$ in the second set of contribution \eqref{secondeoms}, which cures the dependence on higher derivatives. \eqref{secondeoms} is therefore another example of an \eom{} which is healthy, despite the appearance of higher derivative orders. Showing that the decoupling limit interactions of the helicity-0 mode are healthy despite the presence of higher derivative terms is therefore slightly more involved and the galileon dualities discussed in section \ref{sec-dual} will allow us to do so without having to check order-by-order.

\subsection{Multi-Gravity}\label{subsec-multi}

Having covered the massive and bi-gravity cases, we now move on to multi-gravity models. For simplicity we will consider models which do have arbitrarily many spin-2 fields, but where all interaction terms only involve two such fields for any given term. Such setups have an action
\be
\label{MG-act} 
S=\int d^D x  \left[ M_{Pl}^{D-2} \sum_i \sqrt{- g_{(i)}}\,R[g_{(i)}] + \sum_{i,j}^{j > i} \left( m_{(i,j)}^2 M_{\rm Pl}^{D-2} \sqrt{- g_{(i)}}\,\sum_{n=0}^{D} \beta_{(i,j),n}\,
  U_n(\sqrt{g_{(i)}^{-1} g_{(j)}})\right)\right],
\ee
where we have once again assumed all Planck masses to be the same. Note that we have allowed for different coupling constants $m_{(i,j)}$, although we could absorb this into the definition of the coefficients $\beta_{(i,j),n}$, which are now also allowed to vary from interaction term to interaction term. The corresponding decoupling limit action (as before after performing the \St{} trick and projecting out the helicity-0 and -2 components) can be written as a superposition of linearised Einstein-Hilbert terms (EHL) and interaction terms `linking' different spin-2 fields (for more on the link field interpretation see \cite{ArkaniHamed:2001ca,ArkaniHamed:2002sp,ArkaniHamed:2003vb,Schwartz:2003vj,stuck}) as follows\footnote{Technically our prescription for the decoupling limit now becomes slightly more involved. The limit we are taking here is
\begin{align}
m_{(i,j)} &\rightarrow 0, &M_{P} &\rightarrow \infty, &\Lambda_{3,(i,j)} &\text{ fixed},
\end{align}
where $\Lambda_{3,(i,j)}^3 = m_{(i,j)}^2 M_{\rm Pl}$. Note that we could have taken a different decoupling limit, where the coupling constants $m_{(i,j)}$ scale differently and only some of the $\Lambda_{3,(i,j)}$ are kept fixed, while the others tend to zero. The essential point being that there is no unique decoupling limit isolating least suppressed interactions coming from each `link' in the presence of several coupling constants in multi-gravity.}
\bea
{\cal S}_{EHL}^{(i)} &=& -\frac{1}{4} h_{(i)}^{\mu \nu} \hat{\mathcal{E}}^{\alpha \beta}_{\mu\nu} h^{(i)}_{\alpha\beta}, \nn \\
{\cal S}_{link}^{(i,j)} &=& \frac{\Lambda_{3,(i,j)}^3}{2}h^{(i)}_{\mu\nu} X^{\mu\nu}(\Pi_{(i,j)}) + \frac{\Lambda_{3,(i,j)}^3}{2} h^{(j)}_{\mu \nu}\tilde X^{\mu \nu}(\Sigma_{(j,i)}), \nn \\
{\cal S}_{total} &=& \sum_i M_{\rm Pl}^{2} {\cal S}_{EHL}^{(i)} + \sum_{i,j}^{j > i} {\cal S}_{link}^{(i,j)},
\eea
where $(i,j)$ labels refer to the \St{} fields resulting from the interaction term linking $g_{(i)}$ and $g_{(j)}$. As a result the dual to a field labelled by ${(i,j)}$ is labelled by ${(j,i)}$ -- for details see section \ref{subsec-bidual}.
Also note that each link has its own, independent set of coefficients $\hat \beta_{(i,j)}$ implicit in the functions $X$ and $\tilde X$. 
\\

{\bf A concrete trimetric theory}. As a concrete example consider the following trimetric theory
\bea \label{trimetricA}
S &=& \int d^D x \Bigg( M_{Pl}^{D-2} \sqrt{- g_{(1)}}\,R[g_{(1)}] + M_{Pl}^{D-2} \sqrt{- g_{(2)}}R[g_{(2)}] + M_{Pl}^{D-2} \sqrt{- g_{(3)}}\,R[g_{(3)}]  \\
&+&  \left. m^2 M_{Pl}^{D-2} \sqrt{- g_{(1)}}\,\sum_{n=0}^{D} \beta_n\,
  U_n\left(\sqrt{g_{(1)}^{-1} g_{(2)}}\right) + m^2 M_{Pl}^{D-2} \sqrt{- g_{(2)}}\,\sum_{n=0}^{D} \alpha_n\,
  U_n\left(\sqrt{g_{(2)}^{-1} g_{(3)}}\right) \right). \nn
\eea
Here we have three spin-2 fields with Einstein-Hilbert kinetic terms each. There are two bigravity like interaction terms with identical coupling constant $m^2$, which makes taking the decoupling limit trivial, and we have two sets of constant coefficients $\beta_n$ and $\alpha_n$ for the two links respectively (in the above notation these would be $\beta_{n,(1,2)}$ and $\beta_{n,(2,3)}$, but we avoid unnecessary labels as much as possible for the concrete example here.) This theory is depicted in figure \ref{fig-interactions}. The $\Lambda_3$ decoupling limit action now takes on the form
\bea 
S &=& \int \d^Dx \left[  -\frac{1}{4} h_{(1)}^{\mu \nu} \hat{\mathcal{E}}^{\alpha \beta}_{\mu\nu} h^{(1)}_{\alpha\beta} -\frac{1}{4} h_{(2)}^{\mu \nu} \hat{\mathcal{E}}^{\alpha \beta}_{\mu\nu} h^{(2)}_{\alpha\beta} -\frac{1}{4} h_{(3)}^{\mu \nu} \hat{\mathcal{E}}^{\alpha \beta}_{\mu\nu} h^{(3)}_{\alpha\beta} \right.  \nn \\
 &+& \left. \frac{\Lambda_3^3}{2}h^{(1)}_{\mu\nu}(x) X^{\mu\nu}(\Pi_{(1,2)}) + \frac{\Lambda_3^3}{2} h^{(2)}_{\mu \nu}\tilde X^{\mu \nu}(\Sigma_{(2,1)}) \right. \nn \\
&+& \left. \frac{\Lambda_3^3}{2}h^{(2)}_{\mu\nu}X^{\mu\nu}(\Pi_{(2,3)}) + \frac{\Lambda_3^3}{2} h^{(3)}_{\mu \nu}\tilde X^{\mu \nu}(\Sigma_{(3,2)}) \right] \, .\label{MG-dec}
\eea
We now demix this action at linear order via the linearised conformal transformations
\bea
h_{\mn}^{(1)} &\to & h_{\mn}^{(1)} + c_{(1,2)} \pi_{(1,2)} \eta, \nn \\
h_{\mn}^{(2)} &\to & h_{\mn}^{(2)} + c_{(2,1)} \sigma_{(2,1)} \eta + c_{(2,3)} \pi_{(2,3)} \eta, \nn \\
 h_{\mn}^{(3)} &\to & h_{\mn}^{(3)} + c_{(3,2)} \sigma_{(3,2)} \eta, 
\eea
where the $c_{(i,j)}$ are fixed by requiring the tensor-scalar interactions to vanish at linear order. This now leaves us with an action for the pure scalar piece of the schematic form
\bea 
S \sim  \int d^D x \sum_n &\Bigg(& \pi_{(1,2)} U_n (\Pi_{(1,2)}) 
+ \sigma_{(2,1)} U_n (\Sigma_{(2,1)}) + \pi_{(2,3)} U_n (\Sigma_{(2,1)})\nn \\
&+& \sigma_{(2,1)} U_n (\Pi_{(2,3)}) + \pi_{(2,3)} U_n (\Pi_{(2,3)}) +  \sigma_{(3,2)} U_n (\Sigma_{(3,2)})\Bigg), \label{tri-scalar-dec}
\eea
where we have suppressed constant (but $n$-dependent) coefficients for each term and the first and second line derive from the second and third lines in \eqref{MG-dec} respectively. 
\\

\comment{
For a general multi-gravity action built from bimetric links (and {\it{not}} including any loops in its theory graph) we can write down a general schematic form for the decoupling limit action along the same lines  
\bea
S &\sim & \sum_{i,j}^{j \neq i} \Bigg( c_{(i,j)}^{self} \pi_{(i,j)}U_n (\Pi_{(i,j)}) + d_{(i,j)}^{self} \sigma_{(j,i)}U_n (\Sigma_{(j,i)})\Bigg) \nn\\
&+ & \sum_{i,j,k}^{j > i \; ; \; k\neq i,j} \Bigg( c_{(i,j,k)}^{cross} \pi_{(j,i)}U_n (\Sigma_{(j,k)}) + d_{(i,j,k)}^{cross}\sigma_{(j,i)}U_n (\Pi_{(j,k)})\Bigg).
\eea
The relative coefficients of the decoupling limit interactions are a function of the particular coupling constants ($m^2_{(i,j)}, \beta_{n,(i,j)}$) chosen for each bigravity-like interaction term in the original action.
\\
}

{\bf Equations of motion}.
Let us again work out the \eoms{}, this time for the trimetric action \eqref{tri-scalar-dec}. We write $\pi_{(1,2)} = \pi$ and $\pi_{(2,3)}=\phi$ and express the dual fields, $\sigma$ and $\rho$ respectively, in terms of $\pi$ and $\phi$ (the same expression \eqref{sigma-itsolve}, and analogous for $\rho$ in terms of $\phi$, still holds). 
From the bigravity case considered above we already know that $\pi U_n (\Pi) + \sigma U_n (\Sigma)$ and $\phi U_n (\Phi) +  \rho U_n (P)$ give rise to second order Galileon \eoms{} up to quintic order when expressed in terms of $\pi$ and $\phi$. The interesting new piece comes from the scalar mixing generically induced by a multi-gravity theory, i.e. from the terms 
\be
\phi U_n (\Sigma) \; \; \text{and} \; \; \sigma U_n (\Phi).
\ee
It is important to keep in mind that these terms come from $h^{(2)}_{\mu \nu}\tilde X^{\mu \nu}(\Sigma_{(2,1)})$ and $h^{(2)}_{\mu\nu}X^{\mu\nu}(\Pi_{(2,3)})$, so we will give expressions in terms of the associated parameters $\hat\beta_n$ and $\hat\alpha_n$. The contribution of $\phi U_n (\Sigma)$ to the $\pi$ \eom{} up to quartic order in the fields in the action is
\bea
{\cal E}_{(1,\pi)} &=& 0, \nn\\
{\cal E}_{(2,\pi)} &=& \tfrac{1}{2} \hat\beta_1^{\text{}} \phi^{a}{}_{a}, \nn \\
{\cal E}_{(3,\pi)} &=& \hat\beta_1^{\text{}} (\tfrac{1}{2} \phi_{a}{}^{b}{}_{b} \pi^{a} + \tfrac{1}{2} \phi^{a}{}_{a} \pi^{b}{}_{b}) + \hat\beta_2^{\text{}} (\phi^{ab} \pi_{ab} -  \phi^{a}{}_{a} \pi^{b}{}_{b}), \nn \\
{\cal E}_{(4,\pi)} &=& 
\hat\beta_1^{\text{}} (\tfrac{1}{4} \phi_{ab}{}^{c}{}_{c} \pi^{a} \pi^{b} + \tfrac{1}{2} \phi_{a}{}^{c}{}_{c} \pi^{a} \pi^{b}{}_{b} -  \tfrac{1}{4} \phi^{a}{}_{a} \pi_{bc} \pi^{bc} + \tfrac{1}{4} \phi^{a}{}_{a} \pi^{b}{}_{b} \pi^{c}{}_{c}) \nn \\
&+& \hat\beta_2^{\text{}} ( \phi_{abc} \pi^{a} \pi^{bc} - \phi^{ab} \pi_{a}{}^{c} \pi_{bc} -  \phi_{a}{}^{c}{}_{c} \pi^{a} \pi^{b}{}_{b}  + \phi^{a}{}_{a} \pi_{bc} \pi^{bc} + \phi^{ab} \pi_{ab} \pi^{c}{}_{c} -  \phi^{a}{}_{a} \pi^{b}{}_{b} \pi^{c}{}_{c}) \nn \\
&+& \hat\beta_3^{\text{}} (\tfrac{3}{2} \phi^{ab} \pi_{a}{}^{c} \pi_{bc} -  \tfrac{3}{4} \phi^{a}{}_{a} \pi_{bc} \pi^{bc} -  \tfrac{3}{2} \phi^{ab} \pi_{ab} \pi^{c}{}_{c} + \tfrac{3}{4} \phi^{a}{}_{a} \pi^{b}{}_{b} \pi^{c}{}_{c}) 
\eea
and to the $\phi$ \eom{} we find the following contribution
\bea
{\cal E}_{(1,\phi)} &=& 0, \nn \\
{\cal E}_{(2,\phi)} &=& \tfrac{1}{2} \hat\beta_1^{\text{}} \pi^{a}{}_{a} \nn \\
{\cal E}_{(3,\phi)} &=& -\hat\beta_1^{\text{}} (\tfrac{1}{2} \pi_{ab} \pi^{ab} +  \tfrac{1}{2} \pi^{a} \pi_{a}{}^{b}{}_{b}) + \hat\beta_2^{\text{}} (\tfrac{1}{2} \pi_{ab} \pi^{ab} -  \tfrac{1}{2} \pi^{a}{}_{a} \pi^{b}{}_{b}), \nn \\
{\cal E}_{(4,\phi)} &=& 
\hat\beta_1^{\text{}} (\tfrac{1}{2} \pi_{a}{}^{c} \pi^{ab} \pi_{bc} + \pi^{a} \pi^{bc} \pi_{abc} + \tfrac{1}{2} \pi^{a} \pi_{a}{}^{b} \pi_{b}{}^{c}{}_{c} + \tfrac{1}{4} \pi^{a} \pi^{b} \pi_{ab}{}^{c}{}_{c}) \nn \\
&+& \hat\beta_2^{\text{}} (\pi^{a}{}_{a} \pi_{bc} \pi^{bc} - \pi_{a}{}^{c} \pi^{ab} \pi_{bc} -  \pi^{a} \pi^{bc} \pi_{abc} + \pi^{a} \pi^{b}{}_{b} \pi_{a}{}^{c}{}_{c})  \nn \\
&+& \hat\beta_3^{\text{}} (\tfrac{1}{2} \pi_{a}{}^{c} \pi^{ab} \pi_{bc} -  \tfrac{3}{4} \pi^{a}{}_{a} \pi_{bc} \pi^{bc} + \tfrac{1}{4} \pi^{a}{}_{a} \pi^{b}{}_{b} \pi^{c}{}_{c}) 
\eea
From cubic order onwards these are clearly higher-order in derivatives. In this sense the situation is worse than for the bigravity case, where even a disadvantageous choice of demixing procedure only resulted in higher-order derivatives from quartic order in the \eom{} onwards. While we cannot jump to conclusions about the presence of a ghost-like \dof{} here for the same reason as before (the Lagrangian is degenerate due to the non-local way in which $\pi$ and $\phi$ are related to $\sigma$ and $\rho$ respectively), it is far from obvious that there are sufficient constraints in place to prevent extra \dofs{} from propagating. We will need the help of multi-galileon dualities, derived in the following sections, in order to show that this is nevertheless the case.

The contribution of $\sigma U_n (\Phi)$ to the $\pi$ \eom{} is (again up to quartic order in the fields in the action) 
\bea
{\cal E}_{(1,\pi)} &=& 2 \hat\alpha_0^{\text{}}, \nn\\
{\cal E}_{(2,\pi)} &=&  2 \hat\alpha_0^{\text{}} \pi^{a}{}_{a} + \tfrac{3}{2} \hat\alpha_1^{\text{}} \phi^{a}{}_{a}, \nn\\
{\cal E}_{(3,\pi)} &=&  \hat\alpha_0^{\text{}} (\pi^{a}{}_{a} \pi^{b}{}_{b} - \pi_{ab} \pi^{ab})
+ \hat\alpha_1^{\text{}} (\tfrac{3}{2} \phi_{a}{}^{b}{}_{b} \pi^{a} + \tfrac{3}{2} \phi^{a}{}_{a} \pi^{b}{}_{b})
+ \hat\alpha_2^{\text{}} (\tfrac{1}{2} \phi^{a}{}_{a} \phi^{b}{}_{b} - \tfrac{1}{2} \phi_{ab} \phi^{ab}), \nn\\
{\cal E}_{(4,\pi)} &=& 
\hat\alpha_0^{\text{}} (\tfrac{2}{3} \pi_{a}{}^{c} \pi^{ab} \pi_{bc} -  \pi^{a}{}_{a} \pi_{bc} \pi^{bc} + \tfrac{1}{3} \pi^{a}{}_{a} \pi^{b}{}_{b} \pi^{c}{}_{c}) \nn \\ 
&+& \hat\alpha_1^{\text{}} (\tfrac{3}{4} \phi_{ab}{}^{c}{}_{c} \pi^{a} \pi^{b} + \tfrac{3}{2} \phi_{a}{}^{c}{}_{c} \pi^{a} \pi^{b}{}_{b} -  \tfrac{3}{4} \phi^{a}{}_{a} \pi_{bc} \pi^{bc} + \tfrac{3}{4} \phi^{a}{}_{a} \pi^{b}{}_{b} \pi^{c}{}_{c}) \nn \\
&+& \hat\alpha_2^{\text{}} (\phi^{b}{}_{b} \phi_{a}{}^{c}{}_{c} \pi^{a} - \phi^{bc} \phi_{abc} \pi^{a} -  \tfrac{1}{2} \phi_{ab} \phi^{ab} \pi^{c}{}_{c} + \tfrac{1}{2} \phi^{a}{}_{a} \phi^{b}{}_{b} \pi^{c}{}_{c}) \nn \\
&+& \hat\alpha_3^{\text{}} (\tfrac{1}{6} \phi_{a}{}^{c} \phi^{ab} \phi_{bc} -  \tfrac{1}{4} \phi^{a}{}_{a} \phi_{bc} \phi^{bc} + \tfrac{1}{12} \phi^{a}{}_{a} \phi^{b}{}_{b} \phi^{c}{}_{c}), 
\eea
The contribution to the $\phi$ \eom, on the other hand, is
\bea
{\cal E}_{(1,\phi)} &=& 0, \nn\\
{\cal E}_{(2,\phi)} &=& \tfrac{3}{2} \hat\alpha_1^{\text{}} \pi^{a}{}_{a}, \nn\\
{\cal E}_{(3,\phi)} &=& - \hat\alpha_1^{\text{}} (\tfrac{3}{2} \pi_{ab} \pi^{ab} +  \tfrac{3}{2} \pi^{a} \pi_{a}{}^{b}{}_{b}) + \hat\alpha_2^{\text{}} (\phi^{a}{}_{a} \pi^{b}{}_{b} - \phi^{ab} \pi_{ab}), \nn\\
{\cal E}_{(4,\phi)} &=& 
\hat\alpha_1^{\text{}} (\tfrac{3}{2} \pi_{a}{}^{c} \pi^{ab} \pi_{bc} + 3 \pi^{a} \pi^{bc} \pi_{abc} + \tfrac{3}{2} \pi^{a} \pi_{a}{}^{b} \pi_{b}{}^{c}{}_{c} + \tfrac{3}{4} \pi^{a} \pi^{b} \pi_{ab}{}^{c}{}_{c}) \nn \\
&+& \hat\alpha_2^{\text{}} (\phi^{ab} \pi_{a}{}^{c} \pi_{bc} -  \phi^{a}{}_{a} \pi_{bc} \pi^{bc} + \phi^{bc} \pi^{a} \pi_{abc} -  \phi^{b}{}_{b} \pi^{a} \pi_{a}{}^{c}{}_{c}) \nn \\
&+& \hat\alpha_3^{\text{}} (\tfrac{1}{2} \phi_{a}{}^{c} \phi^{ab} \pi_{bc} -  \tfrac{1}{2} \phi^{a}{}_{a} \phi^{bc} \pi_{bc} -  \tfrac{1}{4} \phi_{ab} \phi^{ab} \pi^{c}{}_{c} + \tfrac{1}{4} \phi^{a}{}_{a} \phi^{b}{}_{b} \pi^{c}{}_{c}). 
\eea
Once again we have higher-derivatives dependencies from cubic order onwards and we will have to wait for the multi-galileon dualities, derived in the following sections, in order to show that these interactions are nevertheless healthy.

In summary we see that, unlike for the massive and bigravity cases, the helicity-0 decoupling limit of multi-gravity is not manifestly described by galileon interactions and thus also not manifestly ghost-free. The general proof of \cite{Hassan:2011hr,Hassan:2011ea,Hinterbichler:2012cn} for such models of course ensures that this limit must be ghost-free, but it would immensely help an investigation of the physics of such models to be able to write down the helicity-0 sector in a manifestly second-order form. This is the main aim of the remainder of this paper. In particular we have seen that for multi-gravity models the crucial difference to the bigravity case is the presence of additional cross-terms between different types of `dual' and `original' fields. This is a direct consequence of the `scalar mixing' phenomenon discussed in \cite{stuck}. In the following sections we will find that an extension of the single field galileon dualities \cite{galdual,dualdiff} will allow us to explicitly show that multi-gravity decoupling limit interactions are still of second order form and hence do not propagate any unwanted ghost-like \dofs, linking the multi-gravity decoupling limit to a particular set of multi-galileon-related interactions.

\section{Galileon Dualities}\label{sec-dual}

We have already encountered the \St{} scalar $\pi$ and its `dual' $\sigma$ in the previous sections, which are related by the non-local field redefinition \eqref{sigma-itsolve}. Here we briefly recap the origin of this duality in the bigravity picture \cite{dualdiff}, show that the duality is particularly straightforward to see in the `link field' picture of of bigravity and collect some useful field relations, before extending the duality in the following section to the case of the multi-gravity theories (as introduced in \ref{subsec-multi}).

\subsection{The Bigravity perspective}\label{subsec-bidual}


Consider a generic bigravity theory 
\be \label{bigravity-gen}
S=\int d^D x \bigg[ \sqrt{- g_{(1)}}\,R[{\bf g}_{(1)}] + \sqrt{- g_{(2)}}\,R[{\bf g}_{(2)}] + m^2 V({\bf g}_{(1)},{\bf g}_{(2)})\bigg],
\ee
where $V$ is some potential interaction built out of the spin-2 fields ${\bf g}_{(1)}$, ${\bf g}_{(2)}$ and/or their inverses. Without the potential term this action is invariant under two copies of general co-ordinate invariance (two diffeomorphism symmetries), which we shall call $GC_{(1)}$ and $GC_{(2)}$. The potential term breaks these symmetries down to their diagonal subgroup. The \St{} trick then transforms this action to one with additional fields (the \St{} fields) and more symmetries (the full set of symmetries $GC_{(1)}$ and $GC_{(2)}$ is restored). This symmetry restoration can be understood in the `link field' formulation - for full details in the multi-gravity context we refer to \cite{ArkaniHamed:2001ca,ArkaniHamed:2002sp,ArkaniHamed:2003vb,Schwartz:2003vj,stuck}. 

In the bigravity context there are two possibilities to restore the full diffeomorphism symmetry of the action: Either, via the addition of (gauge) \St{} fields, we transform ${\bf g}_{(1)}$ into an object transforming under $GC_{(2)}$, or we transform ${\bf g}_{(2)}$ into an object transforming under $GC_{(1)}$. This will make all the fields in the potential $V$ transform under the same symmetry group - the potential will no longer break any symmetries, and as a consequence the full symmetry is restored. In terms of the \St{} fields the two `options' are
\begin{align}
g_{\mu\nu}^{(1)} &\to G_{\mu\nu}^{(2)} \equiv \pa_\mu Y^\alpha_{(1,2)} \pa_\nu Y^\beta_{(1,2)} g_{\alpha\beta}^{(1)} &\text{ or }&
&g_{\mu\nu}^{(2)} &\to \tilde G_{\mu\nu}^{(1)} \equiv \pa_\mu \tilde Y^\alpha_{(2,1)} \pa_\nu \tilde Y^\beta_{(2,1)} g_{\alpha\beta}^{(2)},
\end{align}
or, in terms of functional composition notation,
\begin{align}
{\bf g}_{(1)} &\to  {\bf G}_{(2)} \equiv {\bf g}_{(1)} \circ Y_{(1,2)}, &\text{ or }&
&{\bf g}_{(2)} &\to  \tilde {\bf G}_{(1)} \equiv {\bf g}_{(2)} \circ \tilde Y_{(2,1)}.
\end{align}
In terms of the action (we only show the potential term, since the Ricci self-interactions are gauge-invariant and hence are invariant under the \St{} replacement) this means the two possibilities correspond to 
\begin{align}
{\cal S}_{I} &= \int d^D x \sqrt{g_{(1)}} f({\bf g}_{(1)},{\bf g}_{(2)}) \to \int d^D x \sqrt{ g_{(1)} \circ Y_{(1,2)}} f({\bf g}_{(1)} \circ Y_{(1,2)},{\bf g}_{(2)},) \nn \\  &=  \int d^D x \sqrt{G_{(2)}} f({\bf G}_{(2)},{\bf g}_{(2)}),  \\
{\cal S}_{II} &= \int d^D x \sqrt{g_{(1)}} f({\bf g}_{(1)},{\bf g}_{(2)}) \to  \int d^D x \sqrt{ g_{(1)}} f({\bf g}_{(1)},{\bf g}_{(2)} \circ \tilde Y_{(2,1)}) \nn \\ &= \int d^D x \sqrt{g_{(1)}} f({\bf g}_{(1)},{\bf \tilde G}_{(1)}), 
\end{align}
Since the overall action is gauge invariant, we may perform a gauge transformation with parameter $Y_{(1,2)}^{-1}$ on ${\cal S}_{I}^{int}$ or act with $\tilde Y_{(2,1)}^{-1}$ on ${\cal S}_{II}^{int}$, showing that 
\bea
Y_{(1,2)}^{-1} &=& \tilde Y_{(2,1)}, \nn \\
\tilde Y_{(2,1)}^{-1} &=& Y_{(1,2)} \label{Yrelations}
\eea
and we can now extract \St{} scalars describing the helicity-0 mode for each `link field' $Y$ or $\tilde Y$ (where we have implicitly set the helicity-1 modes to zero, as discussed above)
\bea 
Y^{\mu}_{(1,2)} &\to & x^\mu_{(1)} + \pa^\mu \pi_{(1,2)}, \nn \\
\tilde Y^{\mu}_{(2,1)} &\to & x^\mu_{(2)} + \pa^\mu \sigma_{(2,1)}, \label{goldstone1}
\eea
where we have suggestively named the second scalar field $\sigma$. There are consequently two dual and physically equivalent ways of writing down the fully gauge invariant version of a bigravity theory such as \eqref{bigravity-gen}, with the dual scalar fields $\pi$ and $\sigma$. Note that, throughout this paper, we follow the sign convention of \cite{galdual} for the relation between $\pi$ and $\sigma$, which is opposite to that of \cite{gengaldual}.\footnote{These two conventions are related by sending $\sigma \to -\sigma$.} We emphasise that from the link field perspective the existence of this duality is a simple and direct consequence of the fact that there is no unique way to restore gauge invariance in the theory in question. Note that we have kept the labelling of the scalar fields as inherited from the link fields $Y$ and $\tilde Y = Y^{-1}$, so that the dual of $\pi_{(1,2)}$ is $\sigma_{(2,1)}$, i.e. the label indices are reversed.

We may now use \eqref{Yrelations} and \eqref{goldstone1} to implicitly define the {dual field} $\sigma$ in terms of $\pi$ as 
\begin{equation}
(x + \partial \sigma) = (x + \partial \pi)^{-1}. \label{dual relation 2}
\end{equation}
We have already encountered this definition in \eqref{dual relation 1} -- the link formulation now makes it obvious where this definition comes from. One can now solve \eqref{dual relation 2} in order to get an explicit expression for $\sigma$ in terms of $\pi$. This relation can be written succinctly as $Y(Y^{-1}(x))^\mu = x^\mu$. Expanding $Y$ and $Y^{-1}$ in terms of $\pi$ and $\sigma$ respectively we then have
\begin{align}
&(x + \partial \pi)^\mu + \frac{\partial}{\partial(x + \partial\pi)_\mu} \sigma(x + \partial\pi) = x^\mu, \\
\implies \qquad &\partial^\mu \sigma = - \partial^\mu \pi - \sum_{n=1}^{\infty}  D_{(n)} \partial^\mu\sigma, \qquad \text{where} \qquad D_{(n)} = \frac{1}{n!} \pi^{\nu_1} \cdots \pi^{\nu_n} \partial_{\nu_1 \cdots \nu_n}.
\end{align}
This final form allows us to recursively solve for $\partial^\mu \sigma$ in terms of $\partial^\mu \pi$  as\footnote{Note that the first terms in \eqref{sigma-it} are such that $\partial^\mu \sigma$ is a total derivative and so it makes sense to talk of $\sigma$. It is true \cite{dualdiff}, but certainly not obvious (since $[D,\partial] \neq 0$), that this continues to all orders.} 
\begin{equation} \label{sigma-it}
\partial^\mu \sigma = -\partial^\mu( \pi + \frac{1}{2} \pi^\nu \pi_\nu + \dots ).
\end{equation}
In \eqref{sigma-itsolve} we already gave this expansion to quintic order. After some algebra one then reaches an expression for the $n$-th order piece in terms of lower orders ($n$ is an order-label here, {\it not} a space-time index or a field label):
\begin{equation}
\sigma \big|_{\pi^n} = - \sum_{i=1}^{n-1} D_{(n-i)} \left[ \sigma \big|_{\pi^i} \right], 
\end{equation}
which can be solved to give
\begin{equation}
\sigma = - \pi + \sum_{n=2}^\infty \frac{1}{2(n-1)!} \sum_{i=0}^{n-2} (-1)^i \binom{n-2}{i} \tilde{D}_{(i)} \left( \pi^\mu \pi_\mu \mathcal{L}^{\mathrm{TD}}_{(n-2-i)}(\pi) \right), \label{sigmatopi}
\end{equation}
where $\tilde{D}_{(n)}(X) = \partial_{\nu_1 \cdots \nu_n} \left( \pi^{\nu_1} \cdots \pi^{\nu_n} \, X \right)$ for some Lorentz scalar X.
\footnote{We can immediately notice that \eqref{sigmatopi} consists of Galileon terms and total derivatives, so $\sigma$ appearing on its own in the Lagrangian is perfectly healthy when expressed in terms of $\pi$.} We have now explicitly related the dual field $\sigma$ to $\pi$. We again emphasise that in the link field picture the existence of the duality is a direct consequence of gauge invariance and hence is the result of a purely linear phenomenon (in a functional composition sense).

\subsection{Explicit field relations}\label{subsec-fields}

We can abstract the duality transformation away from the link field/multi-gravity argument above and simply view it as a field-dependent diffeomorphism that treats a particular field, e.g. $\pi$, in a privileged way. Here we collect some useful properties of this transformation from \cite{gengaldual} and work out its form for the case of some particular `matter fields', which will turn out to be relevant in section \ref{sec-multidual}. 

We write the duality transformation linked to a field $\pi$ as $\D_\pi$ and now summarise its effect on all the objects in question.\footnote{We introduce the field-dependent label with an eye on the multi-field setups we will deal with in section \ref{sec-multidual}, where there will be a family of field-dependent dualities for each field. Note that this notation is somewhat different from that used in \cite{gengaldual}. \cite{gengaldual} are dealing with a single field galileon duality family, so use of a field label would be redundant, but they do use a further parameter $s$ labelling different types of duality transformations in this setup. We have just taken $s=1$ here for simplicity, absorbing any $s$ into the definition of $\Lambda$, but an explicit generalisation to arbitrary $s$ is straightforward.}
As seen in the bigravity picture above, $\D_\pi$ is a diffeomorphism at heart, so applying $\D_\pi$ leads to a co-ordinate change
\be
\D_\pi: x^\mu  \longrightarrow  \tilde x^{\mu} = x^{\mu} + \frac{1}{\Lambda^{3}}\partial^{\mu} \pi(x),
\ee
where $\Lambda$ is some scale (unsurprisingly we will find that $\Lambda = \Lambda_{3}$ in the bi- and multi-gravity contexts discussed above), which we choose to set to unity in what follows. The \St{} scalar $\pi$ and its derivatives inherit their transformation properties from those of the \St{} fields $Y$, so that in particular the derivative of $\pi$ transforms as a scalar under the action of $\D_\pi$ (since $Y \sim \pa\pi$)
\ba
\label{D1}
\D_\pi: \left\{\begin{array}{rcl}
\pi(x)&\longrightarrow & \sigma(\tilde x) = -\pi( x) - \frac{1}{2} (\partial \pi(x))^2 \, , \\[5pt]
 \p_\mu \pi(x) &\longrightarrow & \tilde \p_\mu \sigma(\tilde x) =  -\p_\mu \pi(x) \,  ,\\[5pt]
\Pi_\mu^\nu (x) &\longrightarrow & \Sigma_\mu^\nu (\tilde x) = -\left[1_\nu^\alpha + \Pi_\nu^\alpha (x) \right]^{-1} \Pi^\alpha_\mu (x), 
\end{array}\right.\,
\ea
where $\Sigma_\mu^\nu (\tilde x) \equiv \tilde \pa_\mu \tilde \pa^\nu \sigma(\tilde x)$,\footnote{
In index-free notation we may also write the expression for $\Sigma$ as $\Sigma = -\left[1 + \Pi \right]^{-1} \Pi$ or equivalently as $\Sigma^{-1} = -(1 + \Pi^{-1})$.} and we have kept the dependence on co-ordinates $x$ and $\tilde x$ explicit, e.g. $\Pi$ is evaluated at $x$, $\Sigma$ at $\tilde x$ here. Note that the transformation properties of $\Pi$ can be derived directly by taking the derivative of the second line in \eqref{D1}. 
When expressed in terms of the same co-ordinates (this amounts to iteratively solving \eqref{D1}) then $\sigma$ and $\pi$ are related non-locally via \eqref{sigma-itsolve}, as discussed. The fact that we can also concisely express their relation in terms of \eqref{D1} shows that this non-locality is a direct consequence of the  (diffeomorphism) co-ordinate transformation mimicked by the \St{} fields. In terms of $\Pi$ (or $\Sigma$) the Jacobian of the co-ordinate transformation $x \to \tilde x$ induced by $\D_\pi$ can now be succinctly expressed as 
\ba
\left|\frac{\partial  x^a}{\partial \tilde x^b}\right|=\det\(1+ \Pi(x)\)^{-1} =\det\(1+ \Sigma(\tilde x)\) \, ,
\ea
assuming that the sign of the determinant is positive.\footnote{For a discussion of this requirement see \cite{galdual}.} As far as fields other than $\pi$ are concerned, $\D_\pi$ is just a diffeomorphism, so a scalar $\chi$, vector $V$ or arbitrary-index tensor $T$ transform just like under diffeomorphisms
\ba \label{DTensor}
\D_\pi: \left\{\begin{array}{rcl}
\chi(x) &\to & \tilde \chi(\tilde x) = \chi(x)\,, \\[5pt]
 V_{\mu}(x) &\to & \tilde V_{\mu}(\tilde x)= \frac{\delta x^\nu}{\delta \tilde x^\mu }V_\nu(x) =
 \left[ 1+ \Sigma(\tilde x) \right]_{\mu}^{\nu}  V_{\nu}(x)\,,\\[5pt]
T_{\mu_1 \dots \mu_n}(x) &\to & \tilde T_{\mu_1  \dots \mu_n}(\tilde x) =
 \left[ 1+ \Sigma(\tilde x) \right]_{\mu_1}^{\nu_1} \cdots\left[ 1+ \Sigma(\tilde x) \right]_{\mu_n}^{\nu_n}T_{\nu_1 \dots \nu_n}(x), \,
\end{array}\right.\,
\ea
Note that we have chosen to write the final expressions in terms of $\Sigma(\tilde x)$, but we could have just as well written them in terms of $\Pi(x)$, e.g. $ \tilde V_{\mu}(\tilde x) = [\left(1+\Pi(x)  \right)^{-1}]^\nu_\mu V_\nu (x)$. The associated inverse duality transformation $\D_\pi^{-1}$ simply involves swapping $\sigma$ and $\pi$ in the above mappings -- for details we refer to \cite{gengaldual}. Here it will turn out to be very useful to work out the explicit action of $\D_\pi$ on another scalar $\chi$ and also on $\pa_\mu \chi$ and $\pa_\mu \pa_\nu \chi$.  
We then have
\ba
\label{Dmatter}
\D_\pi: \left\{\begin{array}{rcl}
\chi(x) &\to & \tilde \chi(\tilde x) = \chi(x)\,, \\[5pt]
 \pa_\mu \chi(x) &\to & \tilde \pa_\mu \tilde \chi (\tilde x) = \frac{\delta x^\nu}{\delta \tilde x^\mu }\pa_\nu \chi(x) =
 \left[ 1+ \Sigma(\tilde x) \right]_{\ \mu}^{\nu}  \pa_\mu \chi(x)\,,\\[5pt]
\pa_\mu \pa_\nu \chi(x) &\to & \tilde \pa_\mu \tilde \pa_\nu \tilde \chi( \tilde x) =
 \left[ 1+ \Sigma(\tilde x) \right]_{\mu}^{\alpha} \pa_\alpha \left( \left[ 1+ \Sigma (\tilde x) \right]_{\nu}^{\beta} \pa_\beta \chi(x) \right)\,.
\end{array}\right.\,
\ea
Throughout the remainder of this paper we will frequently use these maps and talk about `applying' duality maps or duality transformations to an action or \eoms. It is worth emphasising that by this we mean e.g. expressing $V_\nu(x)$ in terms of $\tilde V_\nu (\tilde x)$ by solving the mapping \eqref{DTensor} for $V_\nu(x)$ in terms of $\tilde V_\nu (\tilde x)$. It does {\it not} mean simply replacing $V_\nu(x) \to \tilde V_\nu (\tilde x)$.

\subsection{(Single) Galileon dualities} \label{subsec-singlegal}

A general $n$-th order Galileon interaction term for a single field $\sigma$ may be written
\be
{\cal S}^{\rm Gal}_{(n)}(\sigma) = \int d^D x \sigma U_n(\Sigma).
\ee
If we view $\sigma$ as the dual field for $\pi$, we may now explicitly substitute for $\sigma$ via \eqref{sigmatopi} and, after some algebra, find
\begin{equation}
\int d^D x \sigma U_{(n)}(\Sigma) = \frac{1}{2} (-1)^n (n+1) \int d^D x \left( \sum_{k=0}^{D - n} \frac{(D-n)!}{k!(D-n-k+1)!} \pi^\mu \pi_\mu U_{(n + k -1)}(\pi) + \partial_\mu J^\mu\right),
\end{equation}
which we see is precisely of the form `Galileons + total derivatives' (the exact form of $J_\mu$ is not important here), and upon partial integration is equivalent to
\begin{equation}
\int d^D x \sigma U_{(n)}(\Sigma)  = (-1)^{n+1} (n+1)\int d^D x \left(\sum_{k=n}^{D} \frac{(D-n)!}{(k+1) (D-k)! (k-n)!} \pi U_{(k)}(\pi) + \partial_\mu J^\mu \right). \label{phiLTDphi}
\end{equation}
A single field Galileon in terms of $\sigma$ can therefore be expressed as a Galileon in terms of $\pi$ by direct substitution. In particular this means that a scalar field action as arising in the decoupling limit of Hassan-Rosen bigravity 
\bea
S &=& \int d^D x \sum_{n=0}^{D} c_{(n)} \left(\pi U_{(n)}(\Pi) + \sigma U_{(n)}(\Sigma) \right)\nn\\
&=& \int d^D x \sum_{k=0}^{D} b_{(k)} \pi U_{(k)}(\Pi)
\eea
manifestly describes the \dof{} of a single field Galileon (and is hence ghost-free)\footnote{Note that because $Y(Y^{-1}(x)) = Y^{-1}(Y(x))$, the bigravity picture neatly ensures that these relations take exactly the same form with $\pi$ and $\sigma$ swapped.}, where the coefficients $b_{(n)}$ are given by
\be
b_{(n)} = c_{(n)} +  \sum_{k=0}^{n} \frac{(-1)^{k+1} (k+1)(D-k)!}{(n+1) (D-n)!(n-k)!}c_{(k)}.
\ee
\\

Instead of explicitly substituting for $\sigma$ and subsequently rearranging the action into Galileons plus total derivatives, we may equivalently follow \cite{galdual,gengaldual} and instead use the abstracted duality transformation maps discussed above in \ref{subsec-dmaps} to map a Galileon in one frame into its dual form. Under the duality transformation $\D_\pi$ we then have 
\bea
{\cal S}^{\rm Gal}(\pi) &=& \int d^D x \sum_n c_{(n)} \pi U_{(n)}(\Pi) \\
& \overset{\D_\pi}{\longrightarrow} & -\int \d^D \tilde x \sum_n c_{(n)} \det (1+\Sigma(\tilde x)) \left(\sigma(\tilde x) + \frac{1}{2}\tilde \pa_\mu \sigma(\tilde x) \tilde \pa^\mu \sigma(\tilde x) \right) {U}_{(n)} \left[\frac{-\Sigma(\tilde x)}{1+\Sigma(\tilde x)}\right],\nn
\eea
which after renaming the dummy integration variable $\tilde x$ into $x$ and using some of the algebraic properties of the characteristic polynomial $U_{(n)}$ can be brought into the form
\be
{\cal S}^{\rm Gal}_{(n)}(\pi)  \overset{\D_\pi}{\longrightarrow}       \int \d^Dx\sum_{k=n}^{D} c_{(n)} (-1)^{n+1} \frac{(D-n)!}{(k-n)!(D-k)!} (\sigma + \frac{1}{2}\sigma^\mu \sigma_\mu) {U}_{(k)}[\Sigma(x)]\,,\label{singgaldual}
\ee
where all variables are functions of $x$ now. Now we can already pick out $\sigma U_{(n)}(\Sigma)$ as a Galileon term. For $\sigma^\mu \sigma_\mu U_{(n)}(\Sigma)$ we may integrate by parts to find 
\bea
\int d^D x \sigma_\mu \sigma^\mu U_{(n)}(\Sigma) &=&
(D-n)!\int d^D x \sigma_\mu \sigma^\mu \delta^{\alpha_1 \ldots \alpha_n}_{[\beta_1 \ldots \beta_n]} \sigma^{\beta_1}_{\alpha_1} \ldots \sigma^{\beta_n}_{\alpha_n}\\
&=&
- \frac{2(D-n)!}{n+2}\int d^D x \sigma \delta^{\alpha_1 \ldots \alpha_{n+1}}_{[\beta_1 \ldots \beta_{n+1}]} \pi^{\beta_1}_{\alpha_1} \ldots \pi^{\beta_{n+1}}_{\alpha_{n+1}} \nn \\ &=& -\frac{2(D-n)}{n+2}\int d^D x \sigma U_{(n+1)} (\Sigma) \propto {\cal S}^{Gal}_{(n+1)}(\sigma)\nn.
\eea
Upon substituting this into \eqref{singgaldual}, up to boundary terms we arrive at
\bea
{\cal S}^{\rm Gal}_{(n)}(\pi)   \overset{\D_\pi}{\longrightarrow} (-1)^{n+1}(n+1)\int d^Dx \sum_{k=n}^D  \frac{(D-n)!}{(k+1)(D-k)!(k-n)!} \sigma U_{(k)} (\Sigma),
\eea
which manifestly describes a Galileon in terms of $\sigma$. As expected this has precisely the same form as \eqref{phiLTDphi}.

\section{Multi-Galileon Dualities}\label{sec-multidual}

Knowing how to map single field Galileon interaction terms from one field picture into its dual form, we now move on to multi-Galileon interaction terms. A general such interaction term for $N$ fields at order $n$ in those fields may be written \cite{Padilla:2010ir,Deffayet:2010zh,Hinterbichler:2010xn}
\bea \label{multi-Gal-S}
{\cal S}_{\rm multi-Gal}^{(n)} = \sum_{i_1,\ldots,i_n = 1}^{N} \alpha^{(i_{1},\ldots,i_{n})}\delta^{\nu_2,\ldots,\nu_{n}}_{[\mu_2,\ldots,\mu_{n}]}
\pi_{(i_{1})} {\pi_{(i_{2})}}_{\nu_2}^{\mu_2}\ldots{\pi_{(i_{n})}}_{\nu_n}^{\mu_n},
\eea
where the $(i_j)$ are indices labelling different scalar fields and $\alpha^{(i_1,\ldots,i_{n})}$ are constant coefficients. For the purposes of the remainder of this paper we will restrict ourselves to investigating dualities relevant in the context of the multi-gravity theories discussed in section \ref{subsec-multi} above. As a result two-field Galileons of the following type will be of particular relevance for us
\bea
{\cal S}_{\rm bi-Gal \; I}^{(n)} = \alpha^{(1,2,\ldots,2)}\delta^{\nu_2,\ldots,\nu_{n}}_{[\mu_2,\ldots,\mu_{n}]}
\pi_{(1)} {\pi_{(2)}}_{\nu_2}^{\mu_2}\ldots{\pi_{(2)}}_{\nu_n}^{\mu_n} = \frac{\alpha^{(1,2,\ldots,2)}}{(D-n+1)!}\pi_{(1)}U_{(n-1)}\left(\Pi_{(2)}\right). \label{bi-Gal-S}
\eea
The most general multi-gravity theories naturally will include more general multi-Galileon type interactions in their decoupling limit -- we leave the explicit proof of this to future work.

\subsection{Demixing and the multi-gravity decoupling limit}\label{subsec-dmaps}

Multi-gravity theories as discussed above in section \ref{subsec-multi}, i.e. theories with arbitrarily many spin-2 fields and bigravity-like interaction terms, in the decoupling limit have the following interaction terms (for helicity-2 and -0 modes)
\bea 
{\cal S} &=& - \frac{1}{4}\sum_i h_{(i)}^{\mu \nu} \hat{\mathcal{E}}^{\alpha \beta}_{\mu\nu} h^{(i)}_{\alpha\beta}  + \frac{1}{2}\sum_{i,j}^{j > i} \left( {\Lambda_{3,(i,j)}^3}h^{(i)}_{\mu\nu} X^{\mu\nu}(\Pi_{(i,j)}) + \Lambda_{3,(i,j)}^3 h^{(j)}_{\mu \nu}\tilde X^{\mu \nu}(\Sigma_{(j,i)})\right).
\eea
In order to demix this action at linear order, one can employ the following transformations
\be
h_{\mn}^{(i)} \to  h_{\mn}^{(i)} + \sum_{j}^{}\left(   c_{(i,j)}\pi_{(i,j)} + d_{(i,j)}\sigma_{(i,j)}\right)\eta_{\mn},
\ee
where the constant coefficients $c_{(i,j)}$ and $d_{(i,j)}$ are fixed by the requirement to eliminate mixing between the helicity-2 and -0 modes at linear order, they are in general a function of the particular coupling constants ($m^2_{(i,j)}, \beta_{n,(i,j)}$) chosen for each bigravity-like interaction term in the original action, and we insist on $c_{(i,j)} = 0$, if $d_{(i,j)} \neq 0$ and vice versa\footnote{This choice is always consistent and corresponds to the demixing field re-definition for each $h_{(i)}$ being a function of either $\pi_{(i,j)}$ or $\sigma_{(i,j)}$ for all $j$, but never a function of both.} and $c_{(i,j)} = d_{(i,j)} = 0$ if there is no interaction linking $g_{(i)}$ and $g_{(j)}$ -- for details see \cite{stuck}. As a result, in the demixed decoupling limit, we end up with the pure scalar piece of the action taking the following form
\bea
{\cal S}_{} &\sim & \sum_{i,j}^{j \neq i} \Lambda_{3,(i,j)}^3 \Bigg( c_{(i,j)}   \pi_{(i,j)}U_n (\Pi_{(i,j)}) + d_{(j,i)} \sigma_{(j,i)}U_n (\Sigma_{(j,i)})\Bigg) \nn\\
&+ & \sum_{i,j}^{j \neq i} \sum_{k}^{k \neq i, k \neq j} \Lambda_{3,(i,j)}^3 \Bigg( c_{(j,k)} \pi_{(j,k)}U_n (\Sigma_{(j,i)}) + d_{(i,k)}  \sigma_{(i,k)}U_n (\Pi_{(i,j)})\Bigg)
\nn\\
&+ & \sum_{i,j}^{j \neq i} \sum_{k}^{k \neq i, k \neq j} \Lambda_{3,(i,j)}^3 \Bigg( c_{(i,k)}  \pi_{(i,k)}U_n (\Pi_{(i,j)}) + d_{(j,k)}  \sigma_{(j,k)}U_n (\Sigma_{(j,i)})\Bigg). \label{multidec}
\eea	
The first line describes Galileon-like interactions for $\pi_{(i,j)}$ and the dual fields $\sigma_{(j,i)}$, just as we encountered in the bigravity case. The second and third lines describe new cross-interactions between sets of different $\pi$'s and $\sigma$'s that are qualitatively different. Also note that we have no interaction terms of the type $ \sigma_{(i,k)}U_n (\Pi_{(k,i)})$ and $ \pi_{(i,k)}U_n (\Sigma_{(k,i)})$ -- these are eliminated by choosing a demixing procedure like ${\cal M}_1$ in \eqref{demixings} that yields non-zero $\pi_{(i,j)}U_n (\Pi_{(i,j)})$ and $\sigma_{(j,i)}U_n (\Sigma_{(j,i)})$ interactions, i.e. the first line in \eqref{multidec}. The third line describes cross-interactions, which are already explicitly Bi-Galileons in terms of $\pi$'s or $\sigma$'s. Finally, it is worth pointing out that, for some simple theories, this structure simplifies further, e.g. for line theories (see figure \ref{fig-interactions}) one can make a consistent choice of $c_{(i,j)}$ and $d_{(j,i)}$ that makes the third line in \eqref{multidec} vanish.

In that action we have $N$ fields $\pi_{(i,j)}$ as well as their duals $\sigma_{(j,i)}$. However, since this action describes only $N$ fundamental scalar \dofs, we would like to express the full action (and assess features such as its ghost-freedom, interaction scales etc.) purely in either the $\pi$- or $\sigma$-frame.\footnote{Technically we could choose one field from each $\pi-\sigma$-pair -- which field is labelled $\pi$ and which $\sigma$ for each link is purely conventional.} Here we choose to map the full action into the $\sigma$-frame (this choice is of course completely arbitrary). From section \ref{subsec-bidual} above we know how to map the self-interaction terms from one frame into another and that a single field Galileon in one frame maps into a single field Galileon in the other. However, here we now also have to deal with the new cross-terms, whose mapping we will investigate here. In total we would like to establish the following mappings
\bea
\pi_{(1)} U_{(n)}(\Pi_{(1)}) & \overset{{\D_\pi}_{(1)}}{\longrightarrow} & {\cal A}_{0}(\sigma_{(1)}), \nn\\
\sigma_{(2)} U_{(n)}(\Pi_{(1)}) & \overset{{\D_\pi}_{(1)}}{\longrightarrow} & {\cal A}_{I}(\sigma_{(1)},\sigma_{(2)}), \nn\\
\pi_{(1)} U_{(n)}(\Sigma_{(2)}) & \overset{{\D_\pi}_{(1)}}{\longrightarrow} & {\cal A}_{II}(\sigma_{(1)},\sigma_{(2)}),\nn\\
\pi_{(2)} U_{(n)}(\Pi_{(1)}) & \overset{{\D_\pi}_{(1)}}{\longrightarrow} & \tilde{\cal A}_{III}(\sigma_{(1)},\pi_{(2)})  \overset{{\D_\pi}_{(2)}}{\longrightarrow}  {\cal A}_{III}(\sigma_{(1)},\sigma_{(2)}), \nn\\
\pi_{(1)} U_{(n)}(\Pi_{(2)}) & \overset{{\D_\pi}_{(1)}}{\longrightarrow} & \tilde{\cal A}_{IV}(\sigma_{(1)},\pi_{(2)})   \overset{{\D_\pi}_{(2)}}{\longrightarrow}  {\cal A}_{IV}(\sigma_{(1)},\sigma_{(2)}),  \label{cross-terms}
\eea
figuring out explicit expressions for all the ${\cal A}$ in the process and in particular confirming that the dual formulation of a given term describes precisely the same number of \dofs{} as the term being mapped. In other words we would like to explicitly check that the dynamics of the fields is still described by healthy \eoms{} and consequently that it is explicitly free of ghost-like instabilities. From section \ref{subsec-bidual} we already know that
\be \label{singlegaldual}
{\cal A}_{0}(\sigma_{(1)}) = \sum_n c_{(n)} \sigma_{(1)} U_{(n)}(\Sigma_{(1)}),
\ee
so we will proceed to investigate the cross terms now. Note that $\pi_{(2)} U_{(n)}(\Pi_{(1)})$ and $\pi_{(1)} U_{(n)}(\Pi_{(2)})$ are of course already explicitly Galileons in the $\pi$-frame, but we are here interested in how terms map to the $\sigma$-frame.

\comment{
Taking the results we can already fill in part of our dictionary 
\bea
\pi_{(1)} U_{(n)}(\Pi_{(1)}) & \overset{{\D_\pi}_{(1)}}{\longrightarrow} & \sum_n c_{(n)} \sigma_{(1)} U_{(n)}(\Sigma_{(1)}) \nn\\
\sigma_{(2)} U_{(n)}(\Pi_{(1)}) & \overset{{\D_\pi}_{(1)}}{\longrightarrow} & {\cal A}_{I}(\sigma_{(1)},\sigma_{(2)}) \nn\\
\pi_{(1)} U_{(n)}(\Sigma_{(2)}) & \overset{{\D_\pi}_{(1)}}{\longrightarrow} & {\cal A}_{II}(\sigma_{(1)},\sigma_{(2)}).
\eea
We will now investigate the cross terms.
}

\subsection{Dualities I: Mapping the action}\label{subsec-actmaps}

We begin by investigating the mapping of cross-terms in \eqref{cross-terms} using duality maps at the level of the action, in direct analogy to section \ref{subsec-singlegal}.
\\

 {\bf Bi-Galileon dualities I: } The first term we will consider here is the following cross-interaction term
\bea
\sigma_{(2)} U_{(n)}(\Pi_{(1)}) & \overset{{\D_\pi}_{(1)}}{\longrightarrow} & {\cal A}_{I}(\sigma_{(1)},\sigma_{(2)}). \label{cross-term-one}
\eea
We now aim to employ duality transformations to fully map this into $\sigma$-space at the level of the action, i.e. to  find ${\cal A}_{I}(\sigma_{(1)},\sigma_{(2)})$. Comparison with \eqref{bi-Gal-S} shows that \eqref{cross-term-one} is a two-field Galileon in terms of the two fields $\sigma_{(2)}$ and $\pi_{(1)}$. Here we will show that this is also true when expressed in terms of $\sigma_{(2)}$ and $\sigma_{(1)}$. Taking a general superposition of terms like \eqref{cross-term-one} and applying the duality transformation ${\D_\pi}_{(1)}$ we find 
\bea \label{Scross1}
{\cal S}^{\rm Gal}_{(n)}(\sigma_{(2)},\pi_{(1)}) &=& \int d^D x \sigma_{(2)} U_{(n)}(\Pi_{(1)}) \\
& \overset{{\D_\pi}_{(1)}}{\longrightarrow} & \int \d^D \tilde x \det (1+\Sigma_{(1)}(\tilde x)) \sigma_{(2)} {U}_{(n)} \left[\frac{-\Sigma_{(1)}(\tilde x)}{1+\Sigma_{(1)}(\tilde x)}\right], \nn
\eea
where it is important to note that $\sigma_{(2)}$ transforms like any other `matter' scalar field according to \eqref{Dmatter} under ${\D_\pi}_{(1)}$. As far as $\sigma_{(2)}$ is concerned we are applying a diffeomorphism to the action and, it being a scalar, $\sigma_{(2)}$ remains invariant under that transformation. After renaming the dummy integration variables $\tilde x$ to $x$ and some algebra as before, \eqref{Scross1} can then be brought into the form
\be
{\cal S}^{\rm Gal}_{(n)}(\sigma_{(2)},\pi_{(1)}) \overset{{\D_\pi}_{(1)}}{\longrightarrow}      (-1)^n  \int \d^Dx\sum_{k=n}^{D} \frac{(D-n)!}{(k-n)!(D-k)!} \sigma_{(2)} {U}_{(k)}[\Sigma_{(1)}(x)]\,,\label{bigaction1}
\ee
which is manifestly a two-field Galileon in terms of $\sigma_{(1)}$ and $\sigma_{(2)}$. In other words, for some constant coefficients $\tilde c_{(n)}$ defined via \eqref{bigaction1}, we have
\be
{\cal A}_{I}(\sigma_{(1)},\sigma_{(2)}) = \sum_n \tilde c_{(n)} \sigma_{(2)} U_{(n)}[\Sigma_{(1)}].
\ee
\\

{\bf Bi-Galileon dualities II: } Moving on to the second cross-term in \eqref{cross-terms} we now consider 
\bea
\pi_{(1)} U_{(n)}(\Sigma_{(2)}) & \overset{{\D_\pi}_{(1)}}{\longrightarrow} & {\cal A}_{II}(\sigma_{(1)},\sigma_{(2)}).
\eea      
We could straightforwardly map this term into $\pi$-space using the above recipe, but here we want to uniformly map all possible terms into $\sigma$-space, so we again apply the duality transformations ${\D_\pi}_{(1)}$ and find
\bea \label{Scross2}
 {\cal S}^{\rm Gal}_{(n)}(\pi_{(1)},\sigma_{(2)}) &=& \int d^D x \sum_n c_{(n)} \pi_{(1)} U_{(n)}(\Sigma_{(2)}) \\
& \overset{{\D_\pi}_{(1)}}{\longrightarrow} & -\int \d^D x \sum_n c_{(n)} \det (1+\Sigma_{(1)}) (\sigma_{(1)}+\frac{1}{2}\sigma_{(1)}^\gamma \sigma^{(1)}_\gamma) \nn \\
&\times & {U}_{(n)} \left[ \left[ (1+ \Sigma_{(1)})^{-1} \right]_{\mu}^{\alpha} \pa_\alpha \left( \left[ (1 + \Sigma_{(1)})^{-1}  \right]_{\nu}^{\beta} \sigma^{(2)}_\beta \right)\right]. \nn
\eea
We may now further manipulate this action, e.g. we can extract the factor of $\left[ (1+ \Sigma_{(1)})^{-1}\right]^\alpha_\mu$ from the elementary symmetric polynomial $U_{(n)}$ in order to cancel the determinant factor in front, but the essential feature here is already visible at this stage: This is an action explicitly higher than second order in derivatives. The dependence on $\pa\Sigma_{(1)}$ means we have a dependence on third derivatives of $\sigma_{(1)}$ and we may confirm that the \eoms{} obtained from varying \eqref{Scross2} with respect to $\sigma_{(1)}$ and $\sigma_{(2)}$ are indeed higher order in derivatives as a result -- we will do so below. 

We consequently have a non-Galileon interaction term here and the corresponding \eoms{} will be higher order as well. However, this does not mean that we necessarily have an \Os{} ghost, since as discussed above in section \ref{subsec-higherDeoms} a system can still be secretly second order even though it possesses higher order \eoms{} due to the presence of constraints. This is most easily seen at the level of \eoms, so below we will move on to investigate the multi-Galileon dualities at that level. As expected, this will turn out to be much more straightforward for the second cross-term considered here than mapping at the level of the action.
\\

{\bf Bi-Galileon dualities III: } Before doing so, let us briefly comment on the last two terms in \eqref{cross-terms}. As far as ${\D_\pi}_{(1)}$ is concerned $\pi_{(2)}$ and $\sigma_{(2)}$ transform in exactly the same way, i.e. as a scalar field under diffeomorphisms, so the final two terms in \eqref{cross-terms} transform precisely like the previous two. As such we have (from \eqref{bigaction1}) 
\be
\int d^Dx \sum_n c_{(n)} \pi_{(2)} U_{(n)}[\Pi_{(1)}] \overset{{\D_\pi}_{(1)}}{\longrightarrow}      \int \d^Dx \sum_n \sum_{k=n}^{D} (-1)^n c_{(n)} \frac{(D-n)!}{(k-n)!(D-k)!} \pi_{(2)} {U}_{(k)}[\Sigma_{(1)}(x)]\,,
\ee
and hence 
\bea
\tilde{\cal A}_{III}(\sigma_{(1)},\pi_{(2)}) = \sum_n \tilde c_{(n)} \pi_{(2)} U_{(n)}[\Sigma_{(1)}].
\eea
This can now be fully mapped into the $\sigma$-frame via \eqref{Scross2}, losing its manifest Galileon form in the process as before. For the final cross-term in \eqref{cross-terms} we find (just as in \eqref{Scross2})
\bea 
 \int d^D x \sum_n c_{(n)} \pi_{(1)} U_{(n)}[\Pi_{(2)}] 
& \overset{{\D_\pi}_{(1)}}{\longrightarrow} & -\int \d^D x \sum_n c_{(n)} \det (1+\Sigma_{(1)}) (\sigma_{(1)}+\frac{1}{2}\sigma_{(1)}^\gamma \sigma^{(1)}_\gamma) \nn \\
&\times & {U}_{(n)} \left[ \left[ (1+ \Sigma_{(1)})^{-1} \right]_{\mu}^{\alpha} \pa_\alpha \left( \left[ (1+ \Sigma_{(1)})^{-1}  \right]_{\nu}^{\beta} \pi^{(2)}_\beta \right)\right]. \nn \\\label{Scross3}
\eea
Again this can be fully mapped into the $\sigma$-frame using ${\D_{\pi}}_{(2)}$, which will not, however, restore manifest Galileon form. ${\cal A}_{II}(\sigma_{(1)},\sigma_{(2)})$, ${\cal A}_{III}(\sigma_{(1)},\sigma_{(2)})$ and ${\cal A}_{IV}(\sigma_{(1)},\sigma_{(2)})$ are therefore not of Galileon form and we will now move on to working at the level of the \eoms{} in order to show that, despite their higher-derivative appearance, these dual actions describe healthy theories.

\subsection{Dualities II: Mapping the equations of motion}\label{subsec-eommaps}

We start by considering a generic multi-galileon, whose action is given in \eqref{multi-Gal-S}, which we will here schematically write as
\be
{\cal S}^{\rm Gal}_{(n)}(\pi_{(1)},\pi_{(2)},\ldots,\pi_{(n)})
\ee
The corresponding $i$ \eoms{} that arise from varying this action with respect to each $\pi_{(i)}$ are
\be
{\cal E}_{\pi_{(i)}} \left[ \pi_{(1)}^{II},\pi_{(2)}^{II},\ldots,\pi_{(n)}^{II}  \right] = 0,
\ee
where Roman numerals label the maximal order in derivatives. So, for example, $\pi^{I}$ denotes that the \eom{} depends on first derivatives of $\pi$. In the particular case considered here, multi-galileon \eoms{}, the \eoms{} of course only depend on the $\pi_{(i)}$ via their second-derivatives as is required by the Galilean symmetry (i.e there is no direct dependence on the fields themselves or on first derivatives of the fields). 
\\

{\bf Bi-Galileon duality dictionary: } We now go back to our dictionary of Bi-Galileon terms \eqref{cross-terms}, for which we wanted to establish their mapping under duality transformations and schematically work out their \eoms{} {\it and} how these \eoms{} transform under the duality. 
Our aim is to show that, when completely mapped to the $\sigma$-frame, the resulting $\eoms$ can be put into an explicitly second-order form, demonstrating that no additional ghost-like \dofs{} propagate. If, for a given set of terms, we can show that purely second-order \eoms{} in one frame transform into purely second order \eoms{} in the other frame, then we have successfully shown that Galileons in one field representation map to another set of manifest Galileon interactions under duality transformations  -- as we saw in the previous section this does not have to be true for all types of multi-galileons.\footnote{In other words, multi-galileon \eoms{} in one frame can map into seemingly higher-order \eoms{} in another frame, but still be physically equivalent due to the presence of constraints.} We have
\bea
\pi_{(1)} U_{(n)}(\Pi_{(1)}) & \implies & {\cal E}_{\pi_{(1)}} \left[ \pi_{(1)}^{II} \right] \label{single1}\\
\sigma_{(2)} U_{(n)}(\Pi_{(1)}) & \implies & {\cal E}_{\pi_{(1)}} \left[ \pi_{(1)}^{II},\sigma_{(2)}^{II} \right] \text{ and } {\cal E}_{\sigma_{(2)}} \left[ \pi_{(1)}^{II} \right] \label{cross1}\\
\pi_{(1)} U_{(n)}(\Sigma_{(2)}) & \implies &  {\cal E}_{\sigma_{(2)}} \left[ \pi_{(1)}^{II},\sigma_{(2)}^{II} \right] \text{ and } {\cal E}_{\pi_{(1)}} \left[ \sigma_{(2)}^{II} \right] \label{cross2}\\
\pi_{(2)} U_{(n)}(\Pi_{(1)}) & \implies & {\cal E}_{\pi_{(1)}} \left[ \pi_{(1)}^{II},\pi_{(2)}^{II} \right] \text{ and } {\cal E}_{\pi_{(2)}} \left[ \pi_{(1)}^{II} \right] \label{cross3}\\
\pi_{(1)} U_{(n)}(\Pi_{(2)}) & \implies &  {\cal E}_{\pi_{(2)}} \left[ \pi_{(1)}^{II},\pi_{(2)}^{II} \right] \text{ and } {\cal E}_{\pi_{(1)}} \left[ \pi_{(2)}^{II} \right]. \label{cross4}
\eea
In other words, and again as required for multi-Galileon interactions, the \eoms{} only depend on $\pi_{(i)}$ and $\sigma_{(i)}$ via $\Pi_{(i)}$ and $\Sigma_{(i)}$ respectively. Let us now go through term by term and establish the transformations of these \eoms{} when duality transformations are applied in order to map all fields into the $\sigma$-frame. 
\\

{\bf Galileon dualities: } First the single-field Galileon interactions \eqref{single1}
\be
{\cal E}_{\pi_{(1)}} \left[ \Pi_{(1)} \right] \overset{{\D_\pi}_{(1)}}{\longrightarrow} \hat {\cal E}_{\pi_{(1)}} \left[ -(1+\Sigma_{(1)})^{-1}\Sigma_{(1)} \right] \equiv \hat {\cal E}_{\sigma_{(1)}} \left[\sigma_{(1)}^{II} \right],
\ee
so that the duality transformation manifestly turns a purely second-order equation for $\pi_{(1)}$ into a purely second-order equation for $\sigma_{(1)}$. This proof is of course equivalent to the one via mapping the action shown in section \ref{subsec-actmaps}, which proved that a single Galileon action in terms of $\pi$ straightforwardly maps into one in terms of $\sigma_{(1)}$.
\\

{\bf Bi-Galileon dualities I: } Now for the first cross-term \eqref{cross1} with \eoms{} 
\be \label{bigeom1}
{\cal E}_{\pi_{(1)}} \left[ \pi_{(1)}^{II},\sigma_{(2)}^{II} \right] \text{ and } {\cal E}_{\sigma_{(2)}} \left[ \pi_{(1)}^{II}\right].
\ee
Remembering the transformation properties of $\Pi_{(1)}$ and $\Sigma_{(2)}$ (\eqref{D1} and \eqref{Dmatter} respectively), under $\D_{\pi_{(1)}}$ these \eoms{} are mapped to
\bea
{\cal E}_{\pi_{(1)}} \left[ \pi_{(1)}^{II},\sigma_{(2)}^{II} \right]  &\overset{{\D_\pi}_{(1)}}{\longrightarrow} & \hat{\cal E}_{\pi_{(1)}} \left[ \sigma_{(1)}^{II},\sigma_{(1)}^{III},\sigma_{(2)}^{I},\sigma_{(2)}^{II} \right], \nn \\
{\cal E}_{\sigma_{(2)}} \left[ \pi_{(1)}^{II}\right] &\overset{{\D_\pi}_{(1)}}{\longrightarrow} & \hat {\cal E}_{\sigma_{(2)}} \left[ \sigma_{(1)}^{II}\right]. \label{dualbigeom1}
\eea
These \eoms{} are higher-derivative in $\sigma_{(1)}$ now. However, we may take the second \eom{} and differentiate to obtain 
\be \label{extraeom}
\hat {\cal E}_{\sigma_{(2)}} \left[\sigma_{(1)}^{III}, \sigma_{(1)}^{II}\right] = \pa \hat {\cal E}_{\sigma_{(2)}} \left[ \sigma_{(1)}^{II}\right].
\ee
We can now use $\hat {\cal E}_{\sigma_{(2)}}$ to solve for $\sigma_{(1)}^{III}$ in terms of $\sigma_{(1)}^{II}$ and insert the solution back into \eqref{dualbigeom1}. In that way we finally obtain a set of \eoms{} which are explicitly second-order
\be \label{bi-gal-I-eoms}
\bar {\cal E}_{\pi_{(1)}} \left[ \sigma_{(1)}^{II},\sigma_{(2)}^{I},\sigma_{(2)}^{II} \right] \text{ and } {\cal E}_{\sigma_{(2)}} \left[ \sigma_{(1)}^{II}\right].
\ee
This demonstrates that, in the $\sigma$-frame, a cross-term in the action like \eqref{cross1} does not lead to an Ostrogradsky-ghost via higher-derivatives in the \eoms{} (i.e. signalling a dependence on additional initial conditions). 
Note, however, that in working with the above schematic form for the \eoms, we have not used all of the information available. In particular, while \eqref{bi-gal-I-eoms} shows that the \eoms{} were secretly second order and hence free of any \Os{}-ghost, it does not show that they are purely second-order, i.e. manifestly of the (multi-)Galileon type. We know that this is the case from \eqref{bigaction1}, i.e. $\sigma_{(1)} \text{ and } \sigma_{(2)}$ in fact obey Galilean invariant \eoms{}, so they are indeed bona-fide Bi-Galileons in the $\sigma$-frame. However, without explicitly solving \eqref{extraeom} for $\sigma_{(1)}^{III}$ and carrying out the explicit substitution, the \eom{} analysis carried out here would not show that the dependence on $\sigma_{(2)}^{I}$ in \eqref{bi-gal-I-eoms} vanishes.
\\

{\bf Bi-Galileon dualities II: }
We now look at the \eoms{} coming from the second set of cross-terms \eqref{cross2}
\be \label{crosseoms21}
{\cal E}_{\pi_{(1)}} \left[ \sigma_{(2)}^{II} \right]  \text{ and } {\cal E}_{\sigma_{(2)}} \left[ \pi_{(1)}^{II},\sigma_{(2)}^{II} \right].
\ee
We once again perform the duality transformation $\D_{\pi_{(1)}}$ and find
\bea
{\cal E}_{\pi_{(1)}} \left[ \sigma_{(2)}^{II} \right]  &\overset{{\D_\pi}_{(1)}}{\longrightarrow} & \hat {\cal E}_{\pi_{(1)}} \left[\sigma_{(1)}^{II}, \sigma_{(1)}^{III},\sigma_{(2)}^{II} \right], \nn \\
{\cal E}_{\sigma_{(2)}} \left[ \pi_{(1)}^{II},\sigma_{(2)}^{II} \right] &\overset{{\D_\pi}_{(1)}}{\longrightarrow} & \hat {\cal E}_{\sigma_{(2)}} \left[ \sigma_{(1)}^{II},\sigma_{(1)}^{III},\sigma_{(2)}^{I},\sigma_{(2)}^{II} \right] \label{crosseoms22}
\eea
This time we have a dependence on $\sigma_{(1)}^{III}$ in both \eoms. Now, we could find the linear combination of these \eoms{} that eliminates the third time-derivative of $\sigma_{(1)}$ (and hence, by Lorentz-invariance, also $\sigma_{(1)}^{III}$ as a whole), solve the resulting equation for $\sigma_{(1)}$ (the solution will be given in terms of two initial conditions for $\sigma_{(1)}$ and two for $\sigma_{(2)}$) and then solve one of the original equations for $\sigma_{(2)}$ (since we have already solved for $\sigma_{(1)}$, the fact that there is a third-derivative dependence on  $\sigma_{(1)}$ in the final equation does not introduce extra initial conditions). However, there is no obvious way of eliminating the higher-derivative dependence in \eqref{crosseoms22} and obtaining two equivalent \eoms{} which are explicitly second-order in analogy to \eqref{bi-gal-I-eoms} (at least in the absence of specifying further information about the form of the \eoms{} \eqref{crosseoms22}). 

Nevertheless, there is an alternative way of showing this. We may take \eqref{crosseoms21}, use the second \eom{} to solve for $\sigma_{(2)}$ and insert that solution into the first \eom{}. As a result we will obtain a set of different, yet physically equivalent, differential equations schematically of the form 
\be
\hat{\cal E}_{1} \left[ \pi_{(1)}^{II} \right] \text{ and } \hat{\cal E}_{2} \left[\pi_{(1)}^{II}, \sigma_{(2)}^{II} \right].
\ee
Expressing the dynamical evolution of the system via the equations $\hat{\cal E}_{1},\hat{\cal E}_{2}$ we can now proceed just as for \eqref{bigeom1} above. ${\cal E}_{1},{\cal E}_{2}$ are of precisely the same form as \eqref{bigeom1}, showing that our system secretly obeys second-order \eoms{} in the $\sigma$-frame of the form 
\be \label{bi-gal-II-eoms}
\bar {\cal E}_{\pi_{(1)}} \left[ \sigma_{(1)}^{II},\sigma_{(2)}^{I},\sigma_{(2)}^{II} \right] \text{ and } \hat{\cal E}_{\sigma_{(2)}} \left[ \sigma_{(1)}^{II}\right].
\ee
This explicitly shows that terms in the action like \eqref{cross2} do lead to secretly second-order \eoms{} and hence do not lead to Ostrogradsky ghosts in the $\sigma$-frame. Note, however, that in this dual formulation of the explicitly Galilean-invariant \eoms{} \eqref{crosseoms21}, the Galilean symmetry is not manifest just as in the previous example.
\\

{\bf Bi-Galileon dualities III: } To fully map the remaining terms into the $\sigma$-frame we need to apply both ${\D_\pi}_{(1)}$ and ${\D_\pi}_{(2)}$. Still at the level of the action, using \eqref{bigaction1} we can map
\bea \label{bi-gal-III-I}
\int d^Dx \pi_{(2)} U_{(n)}(\Pi_{(1)}) & \overset{{\D_\pi}_{(1)}}{\longrightarrow} & \sum_n \int d^Dx \tilde c_{(n)} \pi_{(2)} U_{(n)}(\Sigma_{(1)}),\\
\int d^Dx \pi_{(1)} U_{(n)}(\Pi_{(2)}) & \overset{{\D_\pi}_{(2)}}{\longrightarrow} &  \sum_n \int d^Dx \tilde c_{(n)} \pi_{(1)} U_{(n)}(\Sigma_{(2)}),  \label{bi-gal-III-II}
\eea
where the $\tilde c_{(n)}$ are constant coefficients as defined via \eqref{bigaction1}. The \eoms{} from \eqref{bi-gal-III-I} are consequently
\be
\hat{\cal E}_{\pi_{(2)}} \left[ \sigma_{(1)}^{II} \right] \text{ and } \hat{\cal E}_{\sigma_{(1)}} \left[\sigma_{(1)}^{II}, \pi_{(2)}^{II} \right].
\ee
We now apply ${\D_\pi}_{(2)}$ to these \eoms{} and, in precise analogy to the mapping of \eqref{crosseoms21} to the $\sigma$-frame, find that in the $\sigma$-frame all solutions are captured by the following set of second-order differential equations 
\be 
\bar {\cal E}_{1} \left[ \sigma_{(2)}^{II},\sigma_{(1)}^{I},\sigma_{(1)}^{II} \right] \text{ and } \bar{\cal E}_{2} \left[ \sigma_{(2)}^{II}\right].
\ee
The \eoms{} from \eqref{bi-gal-III-II}, on the other hand, are
\be 
{\cal E}_{\sigma_{(2)}} \left[ \sigma_{(2)}^{II},\pi_{(1)}^{II} \right] \text{ and } {\cal E}_{\pi_{(1)}} \left[ \sigma_{(2)}^{II}\right].
\ee
Applying ${\D_\pi}_{(1)}$ to \eqref{bi-gal-III-II} and proceeding as above, in the $\sigma$-frame all solutions are captured by
\be 
\bar {\cal E}_{1} \left[ \sigma_{(1)}^{II},\sigma_{(2)}^{I},\sigma_{(2)}^{II} \right] \text{ and } \bar{\cal E}_{2} \left[ \sigma_{(1)}^{II}\right].
\ee
Again we have shown, at the level of the \eoms, that interactions in the action \eqref{cross3} and \eqref{cross4} do not lead to Ostrogradsky ghosts in the $\sigma$-frame. However, the associated \eoms, which were explicitly of Galileon form in the $\pi$-frame, do not display manifest Galilean symmetry in the $\sigma$-frame (again, just as in the previous example).

\subsection{Dualities III: An explicit multi-galileon example}

The above discussion showed that, while we can use the duality mappings to prove that multi-gravity theories \eqref{MG-act} are free of any (\Os) ghosts in the decoupling limit when mapped into either the $\sigma$- or $\pi$-frame, the resulting interactions and \eoms{} are not always manifestly of Galileon form. While the duality is a direct consequence of diffeomorphism invariance (as the link field picture shows) and the fact that the $\sigma/\pi$-relation is an invertible field re-definition guarantees that the same \dofs{} are described in either frame, it can in principle be difficult to read off symmetries of the helicity-0 decoupling limit interactions.  

For this reason we now briefly give an example of a multi-metric theory in which one can explicitly show that only multi-Galileons appear in the decoupling limit and the second-order nature and a Galilean shift symmetry are therefore manifest, i.e. we do not have the complications of (secretly second-order, but) seemingly higher-order \eoms. Recall the trimetric theory \eqref{trimetricA} and the corresponding demixed decoupling limit action \eqref{tri-scalar-dec}, which we will here schematically write as (suppressing $n$-dependent constant coefficients)
\bea
\mathcal{S} \sim  \int d^D x \sum_n &\Big( & \pi_{(1,2)} U_{n}\left[\Pi_{(1,2)}\right] + \sigma_{(3,2)} U_{n}\left[\Sigma_{(3,2)}\right] \nn \\ &+& (\sigma_{(2,1)} + \pi_{(2,3)}) \left( U_{n}\left[\Sigma_{(2,1)}\right] + U_{n}\left[\Pi_{(2,3)}\right] \right) \Big), \label{trimetric scalar}
\eea
where the first two terms are inherited from mixing with $h_{(1)}$ and $h_{(3)}$ in the pre-demixed action \eqref{MG-dec} respectively (they come from the nodes at the ends of the graph), and terms on the second line are inherited from mixing with $h_{(2)}$ (they come from the middle node). 

From \eqref{singlegaldual} we see that the first two terms, and two arising form the final one are manifestly of Galileon form and hence ghost-free when written either in terms of $\pi_i$ or $\sigma_i$, leaving the two cross-terms terms:
\begin{equation}
\pi_{(2,3)} U_{n}\left(\Sigma_{(2,1)}\right) \qquad \text{and} \qquad \sigma_{(2,1)} U_{n}\left(\Pi_{(2,3)}\right). \label{dangerous terms}
\end{equation}
Simply using \eqref{sigmatopi} in either yields a series of multi-galileon terms as well as a series of additional terms the lowest order of which would be $\partial^a \pi_i \partial_a \pi_{i} \mathcal{L}^\mathrm{TD}_{(n)}(\pi_j)$, which is a Galileon for $i = j$, but leads to higher order equations of motion otherwise.
Now it is if course possible to use the arguments of sections \ref{subsec-actmaps} and \ref{subsec-eommaps} to demonstrate the healthy nature of the whole theory, but in this case it actually possible to approach it in a different way and explicitly show that only multi-Galileons are present.

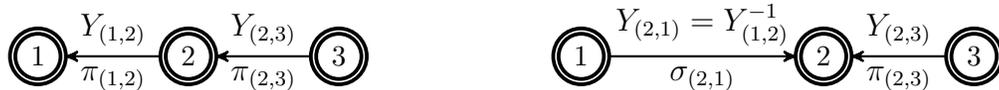
\begin{figure}[tp]
\centering
\begin{tikzpicture}[->,>=stealth',shorten >=0pt,auto,node distance=2cm,
  thick,main node/.style={circle,fill=blue!10,draw,font=\sffamily\large\bfseries},arrow line/.style={thick,-},barrow line/.style={thick,->},no node/.style={plain},rect node/.style={rectangle,fill=blue!10,draw,font=\sffamily\large\bfseries},sarrow line/.style={thick,->,shorten >=1pt},green node/.style={circle,fill=green!20,draw,font=\sffamily\large\bfseries},yellow node/.style={rectangle,fill=yellow!20,draw,font=\sffamily\large\bfseries}]
 
	\node[circle,scale=1.0,fill=white!30,line width=.5mm,draw=black,double] (1) {$1$};
	\node[circle,scale=1.0,fill=white!30,line width=.5mm,draw=black,double] (2) [right of=1] {$2$};
    	\node[circle,scale=1.0,fill=white!30,line width=.5mm,draw=black,double] (3) [right of=2] {$3$};

	\node[circle,scale=1.0,fill=white!30,line width=.5mm,draw=black,double] (4) [right=2.5cm of 3] {$1$};
	\node[circle,scale=1.0,fill=white!30,line width=.5mm,draw=black,double] (5) [right=2.5cm of 4] {$2$};
    	\node[circle,scale=1.0,fill=white!30,line width=.5mm,draw=black,double] (6) [right of=5] {$3$};
  
  	\path[every node/.style={font=\sffamily\large}]
      	(2) edge node [above] {$Y_{(1,2)}$} (1)
	(3) edge node [above] {$Y_{(2,3)}$} (2)
	(2) edge node [below] {$\pi_{(1,2)}$} (1)
	(3) edge node [below] {$\pi_{(2,3)}$} (2)

	(4) edge node [above] {$Y_{(2,1)} = Y_{(1,2)}^{-1}$} (5)
	(6) edge node [above] {$Y_{(2,3)}$} (5)
	(4) edge node [below] {$\sigma_{(2,1)}$} (5)
	(6) edge node [below] {$\pi_{(2,3)}$} (5);

\end{tikzpicture}
\caption{Different orientations of the link fields (as indicated by the arrows, and labels above each edge) lead to taking different fields as the `natural' helicity-0 \dofs{} (indicated below each edge). In order to obtain manifest multi-Galileon interactions for helicity-0 \dofs{} in the trimetric theory considered here, the key if for links to be directed away from \emph{only} nodes of degree one.}\label{fig-linkorientation}
\end{figure}

Looking at the cross terms we immediately note that they only involve $\sigma_{(2,1)}$ and $\pi_{(2,3)}$, and \emph{not} $\pi_{(1,2)}$ or $\sigma_{(3,2)}$, and when written in terms of these fields are manifest bi-Galileons and hence healthy. Thus if we declare that our two `fundamental' helicity-0 degrees of freedom are $\sigma_{(2,1)}$ and $\pi_{(2,3)}$, not $\pi_{(1,2)}$ and $\pi_{(2,3)}$, these terms are now manifestly healthy bi-Galileons. The remaining terms will always be healthy single-field Galileon interactions regardless of which fields are considered `fundamental', where it is worth re-iterating that for each pair of $\pi_{(i,j)}$ and $\sigma_{(j,i)}$, which field is considered to be the `fundamental' one and which the `dual' is purely a matter of convention. Using only single-field dualities we may therefore schematically write \eqref{trimetric scalar} as 
\bea 
S \sim  \int d^D x \sum_n &\Bigg(& \sigma_{(2,1)} U_n [\Sigma_{(2,1)}] + \pi_{(2,3)} U_n [\Pi_{(2,3)}] \nn \\
 &+& \pi_{(2,3)} U_n [\Sigma_{(2,1)}] + \sigma_{(2,1)} U_n [\Pi_{(2,3)}]   \Bigg), \label{tri-scalar-example}
\eea
which as an explicit bi-Galileon interaction for the two scalar helicity-0 \dofs{} $\sigma_{(2,1)}$ and $\pi_{(2,3)}$. In terms of the multi-gravity theory this has a nice interpretation in terms of the direction of the (St\"uckelberg) link fields mapping a metric form one site to another. The original theory has three metrics $g_{(1), (2), (3)}$, and expressing all interactions in the $\pi$-frame amounts to introducing link fields $Y_{(1,2)}$ mapping $g_{(2)}$ onto site $(1)$ in the first interaction term (connecting $g_{(1)}$ and $g_{(2)}$) and $Y_{(2,3)}$ mapping $g_{(3)}$ onto site $(2)$ in the second interaction term (connecting $g_{(2)}$ and $g_{(3)}$), see figure \ref{fig-linkorientation}. Expressing all interactions in the $\sigma$-frame would amount to reversing both links. If, however, we solely change the direction of $Y_{(1,2)}$ so that it instead maps $g_{(1)}$ onto site $(2)$, then we are naturally poised to express things in terms of $\{\sigma_{(2,1)}, \pi_{(2,3)}\}$, again see figure \ref{fig-linkorientation}.

This approach works nicely in the trimetric case considered here, and will continue to work in a theory for which there is only one node of degree greater than one, such as a star graph \cite{stuck,cycles}. However in other cases it fails, due to the fact that terms such as $\sigma_i U_{n} (\Sigma_j)$ are not manifestly of ghost-free Galileon form when written in terms of $\pi_i$ and $\pi_j$. In such cases we must rely on the arguments of sections \ref{subsec-actmaps} and \ref{subsec-eommaps} to demonstrate the healthy second-order nature of the interactions.

\section{Adding a coupling to matter} \label{sec-matter}

Throughout this paper we have dealt with Multi-Gravity theories of the type
\be
{\cal S} = \sum_{i=1}^N M_{Pl}^{D-2} \int d^D x \sqrt{-g_{(i)}} R\left[g_{(i)}\right] +  m^2 M_{Pl}^{D-2} \int d^D x \sqrt{-g_{(1)}} V\left(g_{(1)}, \ldots, g_{(N)}) \right) + {\cal S}_{\rm matter}[\Phi_i,\{g_{(i)}\}], \label{multiS-mat}
\ee
where the matter action ${\cal S}_{\rm matter}[\Phi_i,\{g_{(i)}\}]$, describing the way all matter \dofs{} $\Phi_i$ couple to gravity, has been ignored so far. 
\\

{\bf Minimal coupling and dualities}: The standard, minimal and GR-like, coupling to gravity in Massive, Bi- and Multi-Gravity theories amounts to minimally coupling all matter \dofs{} to a single metric only
\be \label{singlemetriccoupling}
{\cal S}_{\rm matter}[\Phi_i,\{g_{(i)}\}] = {\cal S}_{\rm matter}[\Phi_i,g_{\mn}^{(1)}],
\ee
which we choose to label with the label index $(1)$ here. If the matter metric only couples to one other metric (carrying a label $(2)$), i.e. the matter metric is a node of degree one in the theory graph, then the matter action will only be sensitive to a single helicity-0 mode $\pi$. Expanding around a flat background space-time $\eta_{\mn}$ and focusing on the interactions between that single helicity-0 \dof{} and the matter \dofs{}, from now on for simplicity taken to be described by a single scalar \dof{} $\chi$ (no important feature depends on this assumption, however), we have
\be
{\cal S}_{\rm matter} = \int d^D x {\cal L}_{\rm matter} \left[ \chi(x), \pa_\mu \chi(x), \pi(x), \pa_\mu \pi(x), \pa_\mu\pa^\nu \pi(x)\right],
\ee
where we have assumed that there is no dependence on second derivatives or higher of the matter fields (which is straightforward to generalise via \eqref{Dmatter}, however). Under the duality transformation ${\D}_{\pi_{(1)}}$ this maps to
\bea \label{matterchi}
{\cal S}_{\rm matter} \overset{{\D_\pi}_{(1)}}{\longrightarrow} \int d^D x && \det(1+\Sigma(x)) {\cal L}_{\rm matter} \Big[ \chi(x),  [(1+\Sigma)^{-1}]_\mu^\nu \pa_\nu \chi(x),\\ && -\left(\sigma(x) + \frac{1}{2}(\pa \sigma(x))^2\right), -\pa_\mu \sigma(x), -[(1+\Sigma)^{-1}]_\mu^\nu \pa_\nu \pa^\alpha \sigma(x)\Big], \nn
\eea
where we have replaced the dummy variable $\tilde x$ with $x$  (i.e. after performing the duality transformation). Unlike for single-field Galileons the form of matter interactions is therefore not invariant under the duality (as was also not the case for some of the multi-Galileon interactions considered above). However, both ways of writing down the interaction are of course physically equivalent due to the nature of the duality transformation, and the associated \eoms{} will consequently remain secretly second-order in the sense discussed in the previous sections (also see \cite{gengaldual}).
\\

{\bf Demixing and Decoupling I}: We stay with the case where the matter metric only couples to one other metric (carrying a label $(2)$), as is the case in standard Bi- and Massive Gravity. As part of the demixing procedure we will then perform (up to choices between different consistent demixing procedures, e.g. \eqref{demixings})
\be \label{matter-demix1}
h_{\mn}^{(1)} \to  h_{\mn}^{(1)} + c_{(1,2)}\pi_{(1,2)} \eta_{\mn}.
\ee
As a result we find that the decoupling limit matter action is
\bea
{\cal S}_{\rm matter,dec} &=& \int d^D x \pi(x) T_\mu^\mu \left[ \chi(x), \pa \chi(x) \right],
\eea
where $T_\mu^\mu$ is the trace of the stress-energy tensor defined in the conventional way with respect to metric $g_{(1)}$. This is straightforward to see, if we choose to introduce \St{} fields solely through the metric(s) to which matter does not couple, since the only dependence of the matter coupling on helicity-0 modes then comes in through \eqref{matter-demix1}. Under the duality transformation ${\D}_{\pi_{(1)}}$ this decoupling limit contribution maps to
\bea
{\cal S}_{\rm matter,1,dec} &=& \int d^D x \pi(x) T_\mu^\mu \left[ \chi(x), \pa \chi(x) \right] \\ \nn
&\overset{{\D_\pi}_{(1)}}{\longrightarrow} & -\int d^D x \det(1+\Sigma(x)) \left(\sigma(x) + \frac{1}{2}(\pa \sigma(x))^2\right) T_\alpha^\alpha \left[ \chi(x),  [(1+\Sigma)^{-1}]_\mu^\nu \pa_\nu \chi(x) \right],
\eea
where we have replaced dummy variables as before and we note that, even though the decoupling limit contribution only depends on $\pi$ and not its derivatives, in the dual picture there is explicit dependence on $\sigma,\pa\sigma,\pa^2\sigma$.
\\

{\bf Demixing and Decoupling II}: If the matter metric couples directly to $M$ other metrics (i.e. it is a node of degree $M$ in the theory graph -- see figure \ref{fig-interactions}), then expanding around a flat background space-time $\eta_{\mn}$ and focusing on the interactions between that single helicity-0 \dof{} and the matter \dofs{} as before, we find the following schematic action
\bea
{\cal S}_{\rm matter} = \int d^D x {\cal L}_{\rm matter} \Big[&& \chi(x), \pa_\mu \chi(x), \pi_{(1)}(x), \pa_\mu \pi_{(1)}(x), \pa_\mu\pa^\nu \pi_{(1)}(x), \nn \\ && \hspace{1cm}\ldots,   \pi_{(M)}(x), \pa_\mu \pi_{(M)}(x), \pa_\mu\pa^\nu \pi_{(M)}(x) \Big].
\eea
The demixing procedure for the matter coupling node $(1)$ now takes on the following form (as above we still assume matter only couples to a single metric)
\be
h_{\mn}^{(1)} \to  h_{\mn}^{(1)} + \sum_{j=2}^{M+1}\left(c_{(1,j)}\pi_{(1,j)} + d_{(1,j)}\sigma_{(1,j)}\right)\eta_{\mn},
\ee
leading to a decoupling limit action
\bea
{\cal S}_{\rm matter,dec} &\sim & \int d^D x \sum_{n=1}^{M} \Big(\pi_{(n)}(x)\Big) T_\mu^\mu \left[ \chi(x), \pa \chi(x) \right] \\ \nn
&\overset{{\D_\pi}_{(1)}}{\longrightarrow} & \int d^D x \det(1+\Sigma_{(1)}(x)) \left\{-\left(\sigma_{(1)}(x) + \frac{1}{2}(\pa \sigma_{(1)}(x))^2\right) + \sum_{n=2}^{M} \Big(\pi_{(n)}(x)\Big) \right\} \\ \nn
&& \times \; T_\alpha^\alpha \left[ \chi(x),  [(1+\Sigma_{(1)})^{-1}]_\mu^\nu \pa_\nu \chi(x) \right],
\eea
where we have used a shorthand notation for the helicity-0 fields $\pi_{(1,j)} \equiv \pi_{(j-1)}$ to avoid clutter and ignored constant coefficients throughout. In the same fashion one can now iteratively apply the remaining ${\D_\pi}_{(i)}$ to fully map this into the $\sigma$-frame. Once again the duality guarantees that even if the resulting \eoms{} are higher order, the initial value problem is well-defined in terms of two conditions for each scalar field and hence no extra (\Os) \dofs{} propagate.
\\

{\bf Non-standard matter couplings}: Until now we have only considered matter couplings, in which matter minimally couples to a single metric. However, in general we may have a matter coupling of the form
\be \label{manymetriccoupling}
{\cal S}_{\rm matter}[\Phi_i,\{g_{(i)}\}] = {\cal S}_{\rm matter}[\Phi_i,\tilde g_{\mn}^{\rm matter}[g_{(1)},\ldots,g_{(N)}] ],
\ee
i.e. we couple to more than one metric, where we have insisted that the weak equivalence principle is upheld and matter minimally couples to an effective matter metric $g_{\mn}^{\rm matter}$ that is a function of the other metrics. This can be done consistently (in the decoupling limit) following the matter couplings proposed by \cite{deRham:2014naa,mc}. As a result the matter action, when expanding around a flat background space-time $\eta_{\mn}$ and focusing on the interactions between that single helicity-0 \dof{} and the matter \dofs{} as before, takes on the form
\bea
{\cal S}_{\rm matter} = \int d^D x {\cal L}_{\rm matter} \Big[&& \chi(x), \pa_\mu \chi(x), \pi_{(1)}(x), \pa_\mu \pi_{(1)}(x), \pa_\mu\pa^\nu \pi_{(1)}(x), \nn \\ && \hspace{1cm}\ldots,   \pi_{(N)}(x), \pa_\mu \pi_{(N)}(x), \pa_\mu\pa^\nu \pi_{(N)}(x) \Big].
\eea
Investigating duality mappings in the decoupling limit for these new matter couplings is left for future work. However, one may reasonably expect that these mappings will be rather different from the cases considered above in one important respect. A primary reason why the matter coupling \eqref{singlemetriccoupling} was not invariant under the duality map was that \eqref{singlemetriccoupling} explicitly breaks the symmetry between metrics present at the level of the potential self-interactions. Introducing \St{} fields via one metric or the other in a bigravity-like interaction, while physically equivalent, no longer leaves the form of interactions invariant at the level of the matter action. The many metric coupling \eqref{manymetriccoupling}, especially when considering maximally symmetric matter couplings in multi-gravity theories \cite{mc}, can restore this symmetry and hence potentially some of the duality invariance as we encountered it for single-field Galileons. We leave answering these questions for future work.

\section{Summary and Conclusions}\label{sec-conc}

In this paper we derived the decoupling limit for a particular class of multi-gravity theories and investigated the resulting interactions for helicity-2 and helicity-0 modes. By making use of the link-field picture, we have shown how this is directly related to the existence of dual descriptions for such theories. In particular, we extended the known Galileon dualities to a set of multi-Galileon dualities, relating a class of multi-Galileons to other multi-Galileons and additional higher-derivative interactions, which are healthy due to the existence of constraints. These dualities were discussed in several complementary ways: At the level of the action, the equations of motion, as related to diffeomorphism invariance and as abstracted field re-definitions/mappings. Finally we also showed how matter couplings transform under duality transformations depending on the nature of the coupling. 

Understanding the decoupling limit of multi-Gravity aids us in investigating the physics of these theories and their (low energy) interactions. An obvious task for the future will be to complete the work started here and derive the full decoupling limit interactions for multi-Gravity (i.e. for all ghost-free such models and for all helicity \dofs) and to use this to understand the cosmological phenomenology and relevance of such theories. The dualities uncovered in the process are also of interest from a purely field-theoretic point of view, establishing the equivalence of seemingly unrelated field theories via invertible, non-local field re-definitions that would have been very hard to discover in another way. Realising that these dualities exist can be extremely useful in investigating multi-scalar field models, e.g. for setups where the duality relates strongly and weakly coupled theories and hence makes the physics in seemingly strongly-coupled regimes calculable (this is also the case for the single Galileon duality). It also alerts us to the existence of a rich space of higher-derivative theories, which are nevertheless healthy and free of any ghost-like degrees of freedom. Investigating theories of interacting spin-2 fields appears to be an especially fruitful avenue to discovering healthy theories of this kind.
\\

\noindent {\bf Acknowledgements: } We would like to thank David Alonso, James Bonifacio, Claudia de Rham, Pedro Ferreira, Macarena Lagos, Andrew Matas, Sigurd N\ae ss, Andrew Tolley and Hans Winther for useful discussions and correspondence. JN acknowledges support from the Royal Commission for the Exhibition of 1851, BIPAC and Queen's College, Oxford. JHCS is supported by the STFC.

\appendix

\section{Appendix: Quintic order Lagrangians and quartic order \eoms} \label{appendix-quintic}

In section \ref{sec-dec} we computed the decoupling limit helicity-0 \eom{} for ghost-free Bigravity up to quartic order in the fields in the Lagrangian and hence cubic order in the \eom. Here we complete this task up to quintic order. Extending \eqref{firsteoms}, where the \eom{} was worked out after linear interactions had been demixed with ${\cal M}_1$ from \eqref{demixings}, ${\cal E}_{(5)}$ is given by
\bea
{\cal E}_{(5)} &=& \nn
\hat\beta_1^{\text{}} (- \pi_{a}{}^{c} \pi^{ab} \pi_{b}{}^{d} \pi_{cd} + \tfrac{4}{3} \pi^{a}{}_{a} \pi_{b}{}^{d} \pi^{bc} \pi_{cd} + \tfrac{1}{2} \pi_{ab} \pi^{ab} \pi_{cd} \pi^{cd} -  \pi^{a}{}_{a} \pi^{b}{}_{b} \pi_{cd} \pi^{cd} + \tfrac{1}{6} \pi^{a}{}_{a} \pi^{b}{}_{b} \pi^{c}{}_{c} \pi^{d}{}_{d}) \\ \nn
&+& \hat\beta_2^{\text{}} (\tfrac{9}{2} \pi_{a}{}^{c} \pi^{ab} \pi_{b}{}^{d} \pi_{cd} - 6 \pi^{a}{}_{a} \pi_{b}{}^{d} \pi^{bc} \pi_{cd} -  \tfrac{9}{4} \pi_{ab} \pi^{ab} \pi_{cd} \pi^{cd} + \tfrac{9}{2} \pi^{a}{}_{a} \pi^{b}{}_{b} \pi_{cd} \pi^{cd} -  \tfrac{3}{4} \pi^{a}{}_{a} \pi^{b}{}_{b} \pi^{c}{}_{c} \pi^{d}{}_{d}) \\ \nn
&+& \hat\beta_3^{\text{}} (-6 \pi_{a}{}^{c} \pi^{ab} \pi_{b}{}^{d} \pi_{cd} + 8 \pi^{a}{}_{a} \pi_{b}{}^{d} \pi^{bc} \pi_{cd} + 3 \pi_{ab} \pi^{ab} \pi_{cd} \pi^{cd} - 6 \pi^{a}{}_{a} \pi^{b}{}_{b} \pi_{cd} \pi^{cd} + \pi^{a}{}_{a} \pi^{b}{}_{b} \pi^{c}{}_{c} \pi^{d}{}_{d})\\ \nn
&+& \hat\beta_4^{\text{}} (\tfrac{5}{2} \pi_{a}{}^{c} \pi^{ab} \pi_{b}{}^{d} \pi_{cd} -  \tfrac{10}{3} \pi^{a}{}_{a} \pi_{b}{}^{d} \pi^{bc} \pi_{cd} -  \tfrac{5}{4} \pi_{ab} \pi^{ab} \pi_{cd} \pi^{cd} + \tfrac{5}{2} \pi^{a}{}_{a} \pi^{b}{}_{b} \pi_{cd} \pi^{cd} -  \tfrac{5}{12} \pi^{a}{}_{a} \pi^{b}{}_{b} \pi^{c}{}_{c} \pi^{d}{}_{d}). \\
\eea
Noticeably this is still completely second-order in derivatives, so manifestly free of \Os{} ghosts. Extending \eqref{secondeoms} on the other hand, where demixing had been performed with ${\cal M}_2$ \eqref{demixings}, we find
\bea
{\cal E}_{(5)} &=& \nn
\hat\beta_1^{\text{}} (- \tfrac{1}{2} \pi_{a}{}^{c} \pi^{ab} \pi_{b}{}^{d} \pi_{cd} + \tfrac{1}{6} \pi^{a}{}_{a} \pi_{b}{}^{d} \pi^{bc} \pi_{cd} -  \tfrac{1}{4} \pi^{a}{}_{a} \pi^{b}{}_{b} \pi_{cd} \pi^{cd} + \tfrac{1}{12} \pi^{a}{}_{a} \pi^{b}{}_{b} \pi^{c}{}_{c} \pi^{d}{}_{d} \\ \nn
&-&  \tfrac{3}{2} \pi^{a} \pi_{b}{}^{d} \pi^{bc} \pi_{acd} -  \tfrac{1}{4} \pi^{a} \pi_{bc} \pi^{bc} \pi_{a}{}^{d}{}_{d} + \tfrac{1}{4} \pi^{a} \pi^{b}{}_{b} \pi^{c}{}_{c} \pi_{a}{}^{d}{}_{d} -  \pi^{a} \pi_{a}{}^{b} \pi^{cd} \pi_{bcd} -  \tfrac{1}{2} \pi^{a} \pi^{b} \pi_{a}{}^{cd} \pi_{bcd} \\ \nn
&-&  \tfrac{1}{2} \pi^{a} \pi_{a}{}^{b} \pi_{b}{}^{c} \pi_{c}{}^{d}{}_{d} -  \tfrac{1}{4} \pi^{a} \pi^{b} \pi_{ab}{}^{c} \pi_{c}{}^{d}{}_{d} -  \tfrac{1}{2} \pi^{a} \pi^{b} \pi^{cd} \pi_{abcd} + \tfrac{1}{4} \pi^{a} \pi^{b} \pi^{c}{}_{c} \pi_{ab}{}^{d}{}_{d} -  \tfrac{1}{2} \pi^{a} \pi^{b} \pi_{a}{}^{c} \pi_{bc}{}^{d}{}_{d})\\ \nn
&+& \hat\beta_2^{\text{}} (\tfrac{5}{2} \pi_{a}{}^{c} \pi^{ab} \pi_{b}{}^{d} \pi_{cd} - 3 \pi^{a}{}_{a} \pi_{b}{}^{d} \pi^{bc} \pi_{cd} -  \pi_{ab} \pi^{ab} \pi_{cd} \pi^{cd} + 2 \pi^{a}{}_{a} \pi^{b}{}_{b} \pi_{cd} \pi^{cd} -  \tfrac{1}{2} \pi^{a}{}_{a} \pi^{b}{}_{b} \pi^{c}{}_{c} \pi^{d}{}_{d} \\ \nn
&+& 2 \pi^{a} \pi_{b}{}^{d} \pi^{bc} \pi_{acd} -  \pi^{a} \pi^{b}{}_{b} \pi^{cd} \pi_{acd} -  \pi^{a} \pi^{b}{}_{b} \pi^{c}{}_{c} \pi_{a}{}^{d}{}_{d} + \pi^{a} \pi_{a}{}^{b} \pi^{cd} \pi_{bcd} + \tfrac{1}{2} \pi^{a} \pi^{b} \pi_{a}{}^{cd} \pi_{bcd} \\ \nn &-&  \pi^{a} \pi_{a}{}^{b} \pi^{c}{}_{c} \pi_{b}{}^{d}{}_{d} -  \tfrac{1}{2} \pi^{a} \pi^{b} \pi_{a}{}^{c}{}_{c} \pi_{b}{}^{d}{}_{d} + \pi^{a} \pi^{b} \pi^{cd} \pi_{abcd} -  \pi^{a} \pi^{b} \pi^{c}{}_{c} \pi_{ab}{}^{d}{}_{d}) \\ \nn
&+& \hat\beta_3^{\text{}} (- \tfrac{9}{2} \pi_{a}{}^{c} \pi^{ab} \pi_{b}{}^{d} \pi_{cd} + 6 \pi^{a}{}_{a} \pi_{b}{}^{d} \pi^{bc} \pi_{cd} + \tfrac{9}{4} \pi_{ab} \pi^{ab} \pi_{cd} \pi^{cd} -  \tfrac{9}{2} \pi^{a}{}_{a} \pi^{b}{}_{b} \pi_{cd} \pi^{cd} + \tfrac{3}{4} \pi^{a}{}_{a} \pi^{b}{}_{b} \pi^{c}{}_{c} \pi^{d}{}_{d}) \\ \nn
&+& \hat\beta_4^{\text{}} (\tfrac{5}{2} \pi_{a}{}^{c} \pi^{ab} \pi_{b}{}^{d} \pi_{cd} -  \tfrac{10}{3} \pi^{a}{}_{a} \pi_{b}{}^{d} \pi^{bc} \pi_{cd} -  \tfrac{5}{4} \pi_{ab} \pi^{ab} \pi_{cd} \pi^{cd} + \tfrac{5}{2} \pi^{a}{}_{a} \pi^{b}{}_{b} \pi_{cd} \pi^{cd} -  \tfrac{5}{12} \pi^{a}{}_{a} \pi^{b}{}_{b} \pi^{c}{}_{c} \pi^{d}{}_{d}).  \\
\eea
As had already been the case for ${\cal E}_{(4)}$ for this choice of interactions, there now is explicit higher-order dependence on derivatives, now inside both the $\hat \beta_1$ and $\hat\beta_2$ terms, rather than just inside the $\hat\beta_1$ terms as was the case for ${\cal E}_{(4)}$. Interactions are of course still healthy following the logic of section \ref{subsec-higherDeoms}, but this serves to emphasise that the choice of demixing procedure can be helpful in making the true number of propagating \dofs{} manifest.

\bibliographystyle{JHEP}
\bibliography{paper}

\end{document}